\documentclass[twoside]{article}

\usepackage{qic, epsfig}

\renewcommand{\thefootnote}{\fnsymbol{footnote}}  

\begin{document}
\setlength{\textheight}{8.0truein}    

\runninghead{High-fidelity quantum control using ion crystals in a Penning trap}
            {M. J. Biercuk, H. Uys, A. P. VanDevender, N. Shiga, W. M. Itano, and J. J. Bollinger}

\normalsize\textlineskip
\thispagestyle{empty}
\setcounter{page}{1}

\copyrightheading{0}{0}{2003}{000--000}

\vspace*{0.88truein}

\alphfootnote

\fpage{1}

\centerline{\bf
HIGH-FIDELITY QUANTUM CONTROL}
\vspace*{0.035truein}
\centerline{\bf USING ION CRYSTALS IN A PENNING TRAP}
\vspace*{0.37truein}
\centerline{\footnotesize
MICHAEL J. BIERCUK\footnote{\textit{also} Georgia Institute of Technology, Atlanta, GA, United States},  HERMANN UYS\footnote{\textit{also} Council for Scientific and Industrial Research, Pretoria, South Africa},  AARON P. VANDEVENDER,}
\vspace*{0.015truein}

\centerline{\footnotesize
NOBUYASU SHIGA\footnote{\textit{present address} NICT, Tokyo, Japan},  WAYNE M. ITANO,  AND JOHN J. BOLLINGER}
\vspace*{0.015truein}
\centerline{\footnotesize\it National Institute of Standards and Technology, Time and Frequency Division, 325 Broadway}
\baselineskip=10pt
\centerline{\footnotesize\it Boulder, Colorado 80305, United States}
\vspace*{10pt}

\publisher{(received date)}{(revised date)}

\vspace*{0.21truein}

\abstracts{We provide an introduction to the use of ion crystals in a Penning trap \cite{brel88, Dubin1999, Bollinger2000, Bollinger2003} for experiments in quantum information.  Macroscopic Penning traps allow for the containment of a few to a few million atomic ions whose internal states may be used in quantum information experiments.  Ions are laser Doppler cooled \cite{brel88}, and the mutual Coulomb repulsion of the ions leads to the formation of crystalline arrays \cite{itaw98, mitchell98, jenm04, jenm05}.  The structure and dimensionality of the resulting ion crystals may be tuned using a combination of control laser beams and external potentials  \cite{huap98a, huap98b}.}
{We discuss the use of two-dimensional $^{9}$Be$^{+}$ ion crystals for experimental tests of quantum control techniques.  Our primary qubit is the 124 GHz ground-state electron spin flip transition, which we drive using microwaves \cite{Biercuk2009, BiercukPRA2009}.  An ion crystal represents a spatial ensemble of qubits, but the effects of inhomogeneities across a typical crystal are small,  and as such we treat the ensemble as a single effective spin.  We are able to initialize the qubits in a simple state and perform a projective measurement \cite{brel88} on the system.}
{We demonstrate full control of the qubit Bloch vector, performing arbitrary high-fidelity rotations ($\tau_{\pi}\sim$200 $\mu$s).  Randomized Benchmarking \cite{Knill2008} demonstrates an error per gate (a Pauli-randomized $\pi/2$ and $\pi$ pulse pair) of $8\pm1\times10^{-4}$.  Ramsey interferometry and spin-locking \cite{Redfield1955} measurements are used to elucidate the limits of qubit coherence in the system, yielding a typical free-induction decay coherence time of $T_{2}\sim$2 ms, and a limiting $T_{1\rho}\sim$688 ms.      These experimental specifications make ion crystals in a Penning trap ideal candidates for novel experiments in quantum control.  As such, we briefly describe recent efforts aimed at studying the error-suppressing capabilities of dynamical decoupling pulse sequences \cite{Viola1998,Viola1999, Zanardi1999,Vitali1999, Byrd2003,Khodjasteh2005, Yao2007, Uhrig2007, Uhrig2008, Cywinski2008, Kuopanportti2008, Gordon2008}, demonstrating an ability to extend qubit coherence and suppress phase errors  \cite{Biercuk2009, BiercukPRA2009}.  We conclude with a discussion of future avenues for experimental exploration, including the use of additional nuclear-spin-flip transitions for effective multiqubit protocols, and the potential for Coulomb crystals to form a useful testbed for studies of large-scale entanglement.}

\vspace*{10pt}

\vspace*{3pt}
\communicate{to be filled by the Editorial}

\vspace*{1pt}\textlineskip    
\section{Introduction}
\noindent Trapped ions have proven to be among the leading technologies for the development of quantum information systems.  Typically, experiments are performed using Paul traps which provide confinement to charged particles via a ponderomotive potential arising from an applied radiofrequency field.  A technological alternative to the Paul trap is the Penning trap which provides confinement via the application of static electric and magnetic fields \cite{Major2005}.  These traps have served as the fundamental components of experimental studies in fundamental particle physics \cite{Gabrielse1985, Gabrielse1986, Gabrielse2002, Gabrielse2008}, strongly coupled plasmas (relevant to astrophysics) \cite{Malmberg1977, Horn1991, Dubin1999}, mass spectroscopy \cite{Dyck2004, Marshall2008}, and the engineering of precision frequency standards \cite{Bollinger1994}.  Penning traps, however, have received relatively little attention as an experimental apparatus of choice in quantum information experiments~\cite{thompson2008}.

In this article we describe our ability to implement quantum control techniques with high fidelity using arrays of $^{9}${Be}$^{+}$ ions confined in a macroscopic Penning trap \cite{brel88, Bollinger2000, Bollinger2003}.  We describe the fundamental operational principles and techniques for ion confinement and control in Penning traps \cite{Major2005}, and describe their applicability for experiments in quantum information.  Our experiments employ two-dimensional planar crystals \cite{mitchell98} of $\sim$1000 $^{9}${Be}$^{+}$ ions in a magnetic field of $\sim$4.5 T.  At this large magnetic field we use a ground-state electron-spin-flip transition with a splitting of $\sim$124 GHz as a qubit basis, and implement quantum control via a microwave system \cite{Biercuk2009, BiercukPRA2009}.  We demonstrate through measurements of qubit coherence and characterization of the experimental system that the effects of inhomogeneities are small, and as such we treat the spin ensemble as a single effective qubit.

We present a detailed characterization of measurement fidelity, qubit coherence, and the operational performance of Pauli and Clifford rotations in this system.  Our measurements are based on demonstrations \cite{Vandersypen2004} of high-contrast Rabi flopping, controlled Larmor precession, randomized Clifford benchmarking \cite{Knill2008}, Ramsey free-induction decay, spin-locking \cite{Redfield1955}, and multipulse dynamical decoupling \cite{Viola1998,Viola1999, Zanardi1999,Vitali1999, Byrd2003,Khodjasteh2005, Yao2007, Uhrig2007, Uhrig2008, Cywinski2008, Kuopanportti2008, Gordon2008}.  Experiments reveal long coherence times and high operational fidelity, despite the sensitivity of our qubit transition to magnetic field fluctuations.   In addition to the dominant electron spin-flip transition, we also demonstrate the capability of shelving quantum information in nuclear spin states with $\sim$300 MHz level splitting.  Our control system permits the demonstration of environmental bath engineering \cite{Biercuk2009, BiercukPRA2009}, enabling our qubits to mimic the dynamics of qubit technologies which experience markedly different classical noise environments. In summary, our system may be thought of as a model quantum system, or a precise spin resonance system, providing the ability to perform high-fidelity state initialization and projective measurement.

The remainder of this article is organized as follows.  We begin with an introduction to the basic operating principles of ion storage in a Penning trap in section~\ref{sec:PenningBasics}.  We then move on to a detailed description of our experimental apparatus, including our Penning trap, microwave system, and qubit manifold of choice in section~\ref{sec:NISTPenning}.  Descriptions of high-fidelity quantum control, and detailed system characterizations are provided in sections~\ref{sec:QControl},~\ref{sec:QCoherence} and~\ref{subsec:Nuclear}, followed by a brief discussion of future experiments in section~\ref{sec:Future}.

\section{Penning Trap Basics}\label{sec:PenningBasics}
\noindent  In this section we provide a brief overview of the functional basis for the Penning trap, and highlight some of its salient operational characteristics, largely following the presentation in \cite{Major2005}.  The Penning trap uses static electric and magnetic fields in order to provide confinement for charged particles; an axial, constant magnetic field oriented along the $\hat{Z}$-axis,  $\mathbf{B}=B_{0}\hat{Z}$, and an electric field that provides a harmonic potential well at a locus along this axis.  The ideal trap consists of a central ring electrode that is a hyperbola of revolution and two hyperbolic endcaps, with distance from the center of the trap to the ring $r_{0}$, and distance to the endcap $z_{0}$.  The electrode surfaces are constructed to be equipotential surfaces of the electric potential,
\begin{equation}\label{Eq:Potential}
\Phi=\frac{U_{0}}{2\left(z_{0}^{2}+r_{0}^{2}/2\right)}\left(2z^{2}-x^{2}-y^{2}\right),
\end{equation}
\noindent where $U_{0}$ is the fixed voltage between the ring and end-cap electrodes.   Alternatively, simple cylindrical electrodes may be employed such that the potential is well approximated by Eq.~\ref{Eq:Potential} for distances small relative to the electrode dimensions.

A charged particle is confined axially by the applied electric fields (Fig.~\ref{fig:Trap}a) and radially by the magnetic field.  Writing the particle's equations of motion in three dimensions we find that the particle is axially confined in a harmonic potential with frequency
\begin{equation}
\omega_{z}^{2}=\frac{2qU_{0}}{m\left(z_{0}^{2}+r_{0}^{2}/2\right)}.
\end{equation}
\noindent The radial motion is described by a superposition of two distinct circular motions with frequencies,
\begin{equation}
\omega_{+}=\frac{1}{2}\left(\omega_{c}+\omega_{1}\right),\;\;\;\omega_{-}=\frac{1}{2}\left(\omega_{c}-\omega_{1}\right).
\end{equation}
\noindent Here, $\omega_{c}=qB_{0}/m$ is the cyclotron frequency for charge $q$ and mass $m$, $\omega_{+}$ is known as the modified cyclotron frequency, $\omega_{-}$ is the magnetron frequency, and $\omega_{1}=\sqrt{\omega_{c}^{2}-2\omega_{z}^{2}}$.
Stable confinement is achieved when
\begin{equation}
\omega_{c}^{2}>2\omega_{z}^{2}.
\end{equation}

Thus we see that a charged particle in a Penning trap is harmonically confined in the axial direction, but in the radial plane the particle undergoes complex motion, and traces out what is known as an epitrochoid, due to the high-frequency cyclotron motion and the slower magnetron motion.  The $x-y$ motion can be periodic or quasiperiodic depending upon whether or not the ratio $\omega_{+}/\omega_{-}$ is rational.

\setcounter{footnote}{0}
\renewcommand{\thefootnote}{\alph{footnote}}

\section{The NIST Penning Trap}\label{sec:NISTPenning}
\begin{figure} [htbp]
\centerline{\epsfig{file=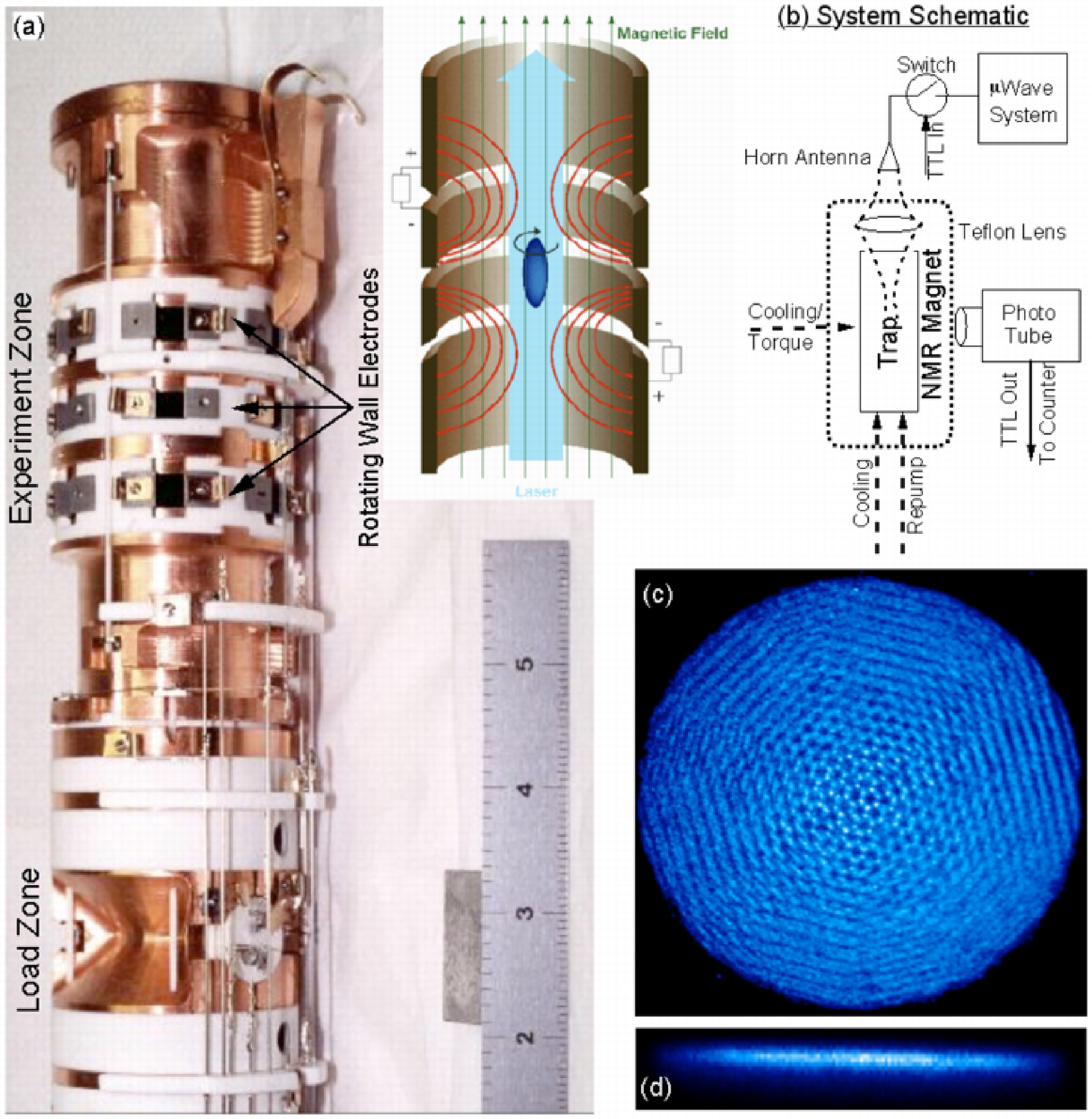, width=12cm}} 
\vspace*{13pt}
\fcaption{\label{fig:Trap} NIST Penning trap.  (a) Photograph of trap electrode structure showing loading trap zone (bottom) and experimental zone (top).  Segmented electrodes in the top half of the image are used for application of a rotating-wall potential.  (Inset) Schematic of experimental zone electrode layout, confining electric fields, ion location (cloud at center), and principal cooling laser direction. (b)Schematic of system providing details of laser access/orientation, imaging system, and microwave system.  Top-view imaging system not shown.  (c-d) Top and side view images of resonant fluorescence from crystallized $^{9}$Be$^{+}$ ion plasmas.  Top view shows hexagonal order, acquired stroboscopically, by gating on the rotating-wall potential.  Multiple ion planes are present, leading to differing degrees of focus among imaged ions.  Side view shows a different crystal with a single plane of ions.}
\end{figure}

\noindent The NIST Penning trap consists of a stack of cylindrical electrodes as pictured in Fig.~\ref{fig:Trap} and described in detail in \cite{brel88, Bollinger2000}.  Charged particles are confined at the symmetry point of the stack along the axial direction, as shown in the inset to Fig.~\ref{fig:Trap}a.  Due to $\mathbf{E}\times\mathbf{B}$ drift from the axial magnetic and radial electric fields, the cloud of charged particles rotates about its axes at a frequency greater than $\omega_{-}$.  Segmented electrodes (Fig.~\ref{fig:Trap}a) are used to apply a rotating dipole or quadrupole potential know as a ``rotating wall.'' Using this external potential it is possible to phase-lock the rotation of trapped ions to the wall, adding stability to the system.

The entire trap structure and enshrouding vacuum envelope is inserted into a 4.5 T room-temperature-bore superconducting magnet with a vertical field orientation. Laser access parallel and perpendicular to the magnetic field axis is provided using apertures and mirrors mounted on the exterior of the glass vacuum envelope. Similarly, optical access for imaging is provided both along the field axis (``Top-view'') and perpendicular to it (``Side-view'').  For the quantum information experiments described in this article, a side-view imaging system using either a spatially resolving camera or a phototube were employed to collect scattered fluorescence from $^{9}${Be}$^{+}$ ions as state-selective readout, as will be described in the next section.  Top-view imaging was also employed in the characterization of ion crystals as will be reviewed in the next section.

\subsection{Trapping and Controlling $^{9}${Be}$^{+}$ Ion Crystals}\label{subsec:TrapControl}
\noindent Previous experimental studies have demonstrated that it is possible to ionize, confine, and laser-cool up to $\sim$$10^{6}$  $^{9}${Be}$^{+}$ atoms in our Penning trap.  Originally, these experiments focused on microwave frequency standards and the physics of cold, one-component plasmas \cite{Bollinger1984, brel88}, appropriate in certain circumstances for studies of dense astrophysical objects.  We present here an overview of the salient physics of these systems that will assist in our understanding of the Penning trap as an experimental testbed for quantum control.

When Doppler laser cooled to sufficiently low temperatures, trapped $^{9}${Be}$^{+}$ ions crystallize due to their mutual Coulomb interaction \cite{itaw98, mitchell98,Bollinger2000}.  This phenomenon is predicted in a model for a classical one-component plasma, and is governed by the Coulomb coupling constant \cite{Ichimaru1987},
\begin{equation}
\Gamma=\frac{1}{4\pi\epsilon_{0}}\frac{e^{2}}{\left(3/4\pi n\right)^{\frac{1}{3}}k_{B}T},
\end{equation}
\noindent where $\epsilon_{0}$ is the permittivity of free space, $e$ is an ion's charge, $k_{B}$ is Boltzmann's constant, $T$ is the ion temperature (which does not include the effects of ion rotation in the confining potential), and $n$ is the plasma density.  The plasma density is set approximately by the physical extent of the ion cloud as limited by the confining potential.  Reducing the temperature relative to the strength of the Coulomb interaction increases $\Gamma$, and above $\Gamma\approx170$ theory predicts that Coulomb repulsion dominates thermal motion and leads to the formation of a body-centered-cubic (BCC) three-dimensional array \cite{Bollinger2000}.

The NIST Penning trap yields an ion cyclotron frequency $\omega_{c}\approx 2\pi \times 7.61$ MHz at 4.46 T, an axial frequency $\omega_{z}\approx 2\pi \times 799$ kHz and a magnetron frequency $\omega_{-}=2\pi \times 42.2$ kHz at a trap voltage of 1 kV.  We have previously demonstrated the transition to BCC order in Doppler cooled $^{9}${Be}$^{+}$ plasmas by both direct imaging \cite{mitchell98, Bollinger2000} and Bragg scattering \cite{itaw98} studies using scattered light from the cooling laser. Calculations indicate that we achieve a coupling constant $\Gamma\approx2000$, for an ion density $n=4\times10^{8}$ cm$^{-3}$, and ion temperature $T\approx1$ mK.  While similar crystallization has been observed in RF Paul traps \cite{Birkl1992}, limitations imposed by the onset of RF heating from trapping fields have prevented the realization of large crystalline arrays with stable BCC ordering.

In thermal equilibrium the ion plasma rotates rigidly in the confining potential.  The rotational frequency is controllable via both an external rotating potential \cite{huap98a, huap98b} and a cooling laser beam applied perpendicular \cite{brel88} to $\mathbf{B}$.  For example,  the perpendicular laser induces a torque on the plasma due to radiation pressure, which can be used to increase or decrease the rotational frequency.  The rotation frequency, $\omega_{r}$, controls the shape and dimensionality of the ion plasma.  This dependence arises from the fact that the radial confinement associated with ion motion in the magnetic field depends on the rotational velocity \cite{brel88, Bollinger2003}.  Thus by changing $\omega_{r}$ it is possible to adjust the effective radial trapping potential.  The ion plasma adjusts its physical extent to the strength of the radial confinement: for weak confinement near the lower bound $\omega_{r}\approx\omega_{-}$ the cloud assumes an oblate form, while for $\omega\approx\omega_{c}/2$ the cloud spins up to an oblong cigar-like shape along the magnetic field axis.  As the rotational frequency increases up to $\omega_{r}\approx\omega_{+}$, the confining centripetal acceleration becomes insufficiently strong to maintain the axially extended shape of the ions and the cloud relaxes towards an oblate shape.  At higher rotational frequencies the balance of the outwardly directed radial electric force and the confining Lorentz force does not provide confinement and the radial extent of the cloud diverges.  For the quantum information experiments described in this work we focus on oblate clouds at low rotation frequencies with $\sim$100-1000 ions in one or two planes, easily resolvable via side-view imaging (Fig.~\ref{fig:Trap}d), and having radial extent of a few hundred microns.

Once the rotational frequency is coarsely set via the perpendicular laser, the ions are locked to an external dipole ``rotating wall'' potential \cite{huap98a, huap98b}, providing long-term stability over many hours.  Figure~\ref{fig:Trap}c shows a top view image of an ion array obtained with a long integration time ($\sim$30 s).  This image was obtained with an intensified CCD camera, strobing the camera synchronously with the applied rotating potential used to control the plasma rotation frequency.  The figure is therefore an image of the ions in the frame of the rotating wall potential.  Ions are well resolved in the array center, but not near the edges of the array. As discussed in \cite{Mitchell2001} this difference is due to small slips of the planar crystal relative to the rotating wall potential. Studies of the ``stick-slip'' dynamics of these crystals~\cite{mitchell98, Mitchell2001} have revealed that the ions experience abrupt reorientations of the crystal structure in the frame of the rotating wall, separated by stable phase-locked rotation for periods many tens of seconds long.

The dominant ion loss mechanisms limiting the ion cloud lifetime are background collisions in which $^{9}${Be}$^{+}$ is converted to BeH$^{+}$ or BeOH$^{+}$ \cite{Bollinger2003}.  These molecular ions remain part of the crystallized plasma, and are sympathetically cooled by the $^{9}${Be}$^{+}$ ions.  These so-called ``heavy-mass'' ions centrifugally separate from the lighter $^{9}${Be}$^{+}$, and due to their modified level structures do not significantly participate in any quantum control experiments.  Ion collisions with background gas molecules typically reduce the number of $^{9}${Be}$^{+}$ ions by half over a timescale of one to two days.

\subsection{$^{9}${Be}$^{+}$ Level Structure at 4.5 T}
\begin{figure} [htbp]
\centerline{\epsfig{file=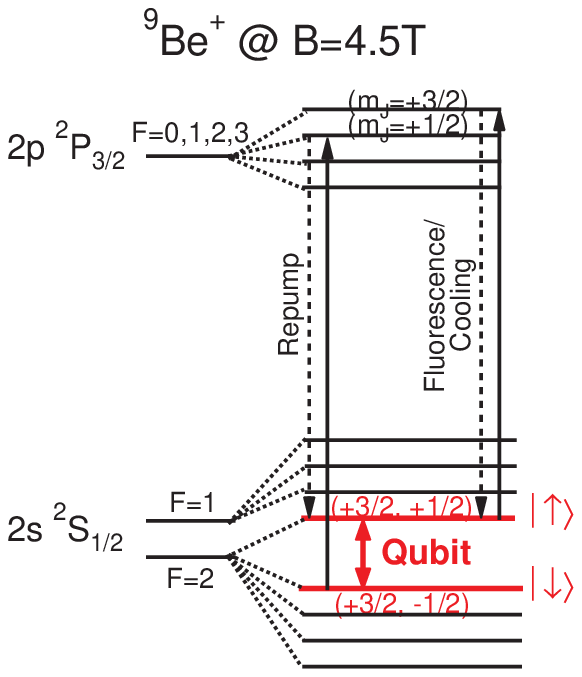, width=8cm}} 
\vspace*{13pt}
\fcaption{\label{fig:Levels}  Level structure of $^{9}$Be$^{+}$ at 4.5 T.  Highlighted levels correspond to electron-spin-flip transition used as primary qubit basis.  Numbers in parentheses represent ($m_{I},m_{J}$), the nuclear and electron spin projections along the quantization axis. The upper qubit state is purely defined by these quantum numbers while other states in the hyperfine manifold contain a small amount of admixing of other ($m_{I},m_{J}$) states ($m_{F}$ is a good quantum number).  Repump and Fluorescence/Cooling transitions are near 313 nm.  Qubit splitting is near 124 GHz.}
\end{figure}

\noindent The level structure of $^{9}${Be}$^{+}$ ions is modified by the presence of an external confining magnetic field with strength $\sim$4.5 T.  The relevant energy levels are depicted schematically in Fig.~\ref{fig:Levels} \cite{brel88}.  Our primary qubit states are defined by two states with 124 GHz separation within the $2s^{2}S_{1/2}$ manifold,
\begin{equation}
\left|F=2, m_{I}=+3/2, m_{J}=+1/2\right\rangle\equiv\left|\uparrow\right\rangle\leftrightarrow\left|F=2, m_{I}=+3/2, m_{J}=-1/2\right\rangle\equiv\left|\downarrow\right\rangle,
 \end{equation}
 where $m_{I}$ and $m_{J}$ are the nuclear and electron spin projections along the quantizing field, and $F$ is the hyperfine quantum number.  Around 4.5 T, the qubit states diverge linearly at 28 MHz/mT, making this transition sensitive to magnetic field fluctuations.  The ``bright'' $\left|\uparrow\right\rangle$ state is coupled to the $2p^{2}P_{3/2}(m_{J}=+3/2)$ manifold by a cycling transition at 313 nm.  This transition is employed for Doppler cooling (when red-detuned), plasma control (via radiation pressure in a perpendicular configuration) and state-selective readout, since $\left|\uparrow\right\rangle$ is optically bright.  The ``dark'' state $\left|\downarrow\right\rangle$ is minimally affected by the radiation on the cycling transition~\fnm{a}\fnt{a}{The dark state is repumped to the bright state by the $\sim$313 nm cooling light on a timescale of 5-10 ms.}, but can be rapidly repumped by another transition near 313 nm to the $2p^{2}P_{3/2}(m_{J}=+1/2)$ manifold.

\subsection{Microwave and RF Control Systems}
\begin{figure} [htbp]
\vspace*{13pt}
\centerline{\epsfig{file=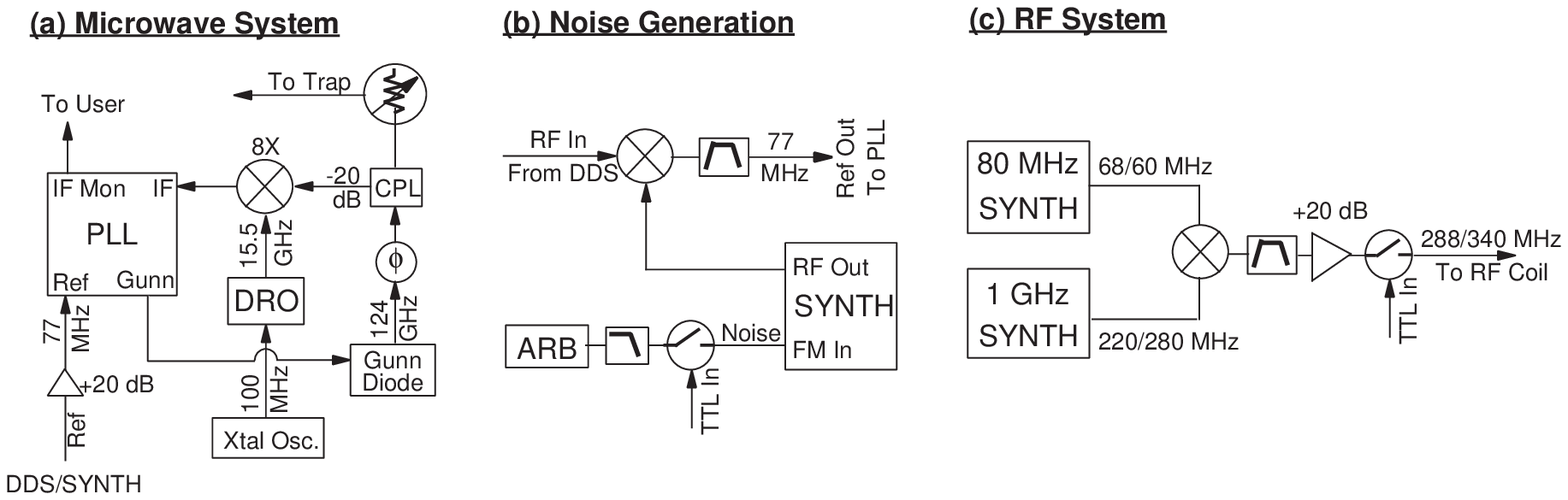, width=12cm}} 
\vspace*{13pt}
\fcaption{\label{fig:Electronics} Schematic block diagrams of a) 124 GHz microwave system, b) Noise generation system, and c) RF system. PLL = Phase Lock Loop, DRO = Dielectric Resonator Oscillator, CPL = Directional Coupler, DDS = Direct Digital Synthesizer, ARB = Arbitrary waveform generator, SYNTH = frequency synthesizer, TTL = Transistor-Transistor Logic pulse.}
\vspace*{13pt}
\vspace*{13pt}
\end{figure}

\noindent Transitions between the qubit states are driven by a home-built microwave system \cite{Biercuk2009, BiercukPRA2009} as depicted in Fig.~\ref{fig:Electronics}.  The system is based on a Gunn diode oscillator which outputs approximately 30 mW at 124 GHz.  This oscillator is phase-locked to the sum of the eighth harmonic of a 15.5 GHz DRO and an external reference at $\sim$77 MHz (for definitions of hardware abbreviations see Fig.~\ref{fig:Electronics}).

The DRO employs a 100 MHz quartz crystal oscillator as an ultra-stable, low-phase-noise reference.  The bandwidths of the DRO PLL (to the 100 MHz crystal oscillator) and the 124 GHz Gunn diode oscillator PLL (to the DRO) are 100 kHz and several times 100 kHz, respectively.  Accordingly, the phase noise of the microwave source for frequency offsets less than 100 kHz is determined by the phase noise of the 100 MHz crystal oscillator.  Our 100 MHz crystal oscillator has a maximum phase noise specification of -125 dBc/Hz, -150 dBc/Hz, -175 dBc/Hz, and -175 dBc/Hz at 100 Hz, 1 kHz, 10 kHz, and 100 kHz respectively.  This implies a maximum integrated phase noise of $\sim$107 dBc over the experimentally relevant interval $[100\;\rm{Hz}, 100\;\rm{kHz}]$.  The phase noise of the microwaves is increased by $20\log(N)$ = 62 dB where $N$=1240 is the step-up factor from the 100 MHz oscillator to the 124 GHz microwaves, leading to an integrated phase noise of $\sim-$45 dBc at the 124 GHz carrier over the same interval.

The $\sim$77 MHz reference is provided by a computer-controlled direct digital synthesizer (DDS) which is referenced to a 1 GHz clock derived from the 5 MHz signal of a passive hydrogen maser.  Changes in the $\sim$77 MHz signal are tracked linearly by the Gunn Diode, making this reference the primary means of tuning the output of the microwave system.  The DDS output may be switched between four pre-programmed output profiles via TTL inputs, each with a user-defined frequency and phase, and the Gunn diode PLL tracks changes in DDS output within $\sim$5 $\mu$s.

The configuration described above is engineered for maximum amplitude, phase, and frequency stability.  For quantum control experiments, however, it is occasionally desirable to engineer the noise environment to exhibit precisely defined spectral characteristics, as will be seen in section~\ref{sec:DD}.  As it is difficult to inject magnetic field noise into a superconducting magnet system, we instead engineer noise in the microwave control system \cite{Biercuk2009, BiercukPRA2009}, which in the appropriate frame of reference is indistinguishable from the action of classical magnetic field fluctuations on the qubit states.  We replace the $\sim$77 MHz reference for the Gunn diode PLL with a frequency-modulated (FM) signal generated by a synthesizer (Fig.~\ref{fig:Electronics}b).  A desired spectrum is generated numerically and Fourier transformed to produce a time-domain data set whose two-time correlation function reproduces the spectrum of interest.  These data are sent to an arbitrary waveform generator whose voltage output is sent to the FM-input of the synthesizer, producing a carrier at 275 MHz with the desired noise characteristics. This carrier is mixed with a tunable $\sim$197 MHz output from the computer-controlled DDS, band-pass filtered, and sent to the Gunn diode PLL, which linearly tracks the frequency-modulated reference to a bandwidth of $\sim$100 kHz.

Microwave power is output from the Gunn diode to a horn antenna via rigid waveguides.  Power reaching the antenna may be tuned via a manual attenuator, and pulsed-control is afforded by a p-i-n diode switch that provides $\sim$25 dB of isolation in the off-state.  Using this hardware configuration, the only available pulse shape is approximately square, and the relatively long lengths of the relevant pulses ($\sim$100-200 $\mu$s) yield a nearly ideal square envelope.  Output from the microwave horn propagates in free-space and is focused on the ion crystal using a custom teflon lens installed in the bore of the superconducting magnet.  The microwave beam waist at the position of the ions is approximately 7 mm, allowing for large incident microwave power on the ions.  The beam waist and microwave wavelength are large relative to the extent of the ion crystal, allowing for high driving-field homogeneity.

An RF coil antenna is located outside of the vacuum envelope, in close proximity to the location of the ions, and is used to drive nuclear-spin-flip transitions as will be discussed in section~\ref{subsec:Nuclear}.  Coarse and fine synthesizer outputs are mixed and bandpass filtered to provide the appropriate carrier frequencies with resolution of $\sim$1 mHz (Fig.~\ref{fig:Electronics}c).  Both synthesizers are phase-locked to the output of a passive hydrogen maser, as used for the DDS.  The carrier signal is amplified using a standard RF amplifier and power reaching the antenna  is controlled via a commercial RF switch under TTL control.

Pulsed-control over the qubit drive and RF systems is achieved via a programmable-logic device.  Details of the experimental control and pulse-sequencing systems,  as well as the user interface, are provided in reference~\cite{BiercukPRA2009}.

\section{Quantum Control Experiments}\label{sec:QControl}
\noindent All experimental sequences begin with optical pumping~\fnm{a}\fnt{a}{Optical pumping of the nuclear spins occurs due to mixing of different  $\left | m_{I}, m_{J} \right \rangle$ states in the excited $P$-state manifold over a time $\sim1$ s. For quantum control experiments using the 124 GHz electron-spin-flip transition this repumping does not need to be repeated because the nuclear spin does not leave the $m_{I}=+3/2$ manifold.} of the ions to the bright state via the cooling/detection and repump lasers near 313 nm.  Pulsed exposure for $\sim$100 ms, synchronous with line frequency initializes the ions to $\left|\uparrow\right\rangle$ with infidelity $\sim1\times10^{-6}$, estimated from the scatter rate of the repumping laser and the $\left|\uparrow\right\rangle$ to $\left|\downarrow\right\rangle$ transition rate due to microwave leakage.  Ion fluorescence is integrated over 50 ms using an $f/5$ optical system on a phototube, an experimental microwave or RF pulse sequence is executed, and ion fluorescence is again measured for 50 ms.  In all experiments, the ions are rotated to the dark state, $\left|\downarrow\right\rangle$, at the end of the control sequence. The second measurement cycle is broken into five 10 ms bins and a linear fit to the count rate provides the fluorescence level at the end of the experimental control sequence.  This technique accounts for repumping due to the detection laser from $\left|\downarrow\right\rangle$ to $\left|\uparrow\right\rangle$.   The extracted fluorescence count rate is normalized to the bright-state count rate measured at the beginning of the experimental cycle to mitigate the effects of slow fluctuations in laser amplitude between experimental runs.  Experimental data are generally presented as a normalized fluorescence count rate, which is commensurate with population in $\left|\uparrow\right\rangle$, and is sometimes used as a measure of the qubit coherence, as will be described in the context of each experiment.  Each data point presented in an experiment typically consists of an average of 20-50 individual experiments, all performed under approximately the same conditions.

In the following subsections we will describe a series of quantum control experiments, primarily using the electron-spin-flip qubit basis.  Each experiment consists of an appropriate microwave or RF pulse chain applied in the experimental sequence described above.  These experiments will elucidate our quantum control capabilities, and will highlight the strengths and limitations of this experimental system.

\subsection{Driven Rotations}
\begin{figure} [htbp]
\centerline{\epsfig{file=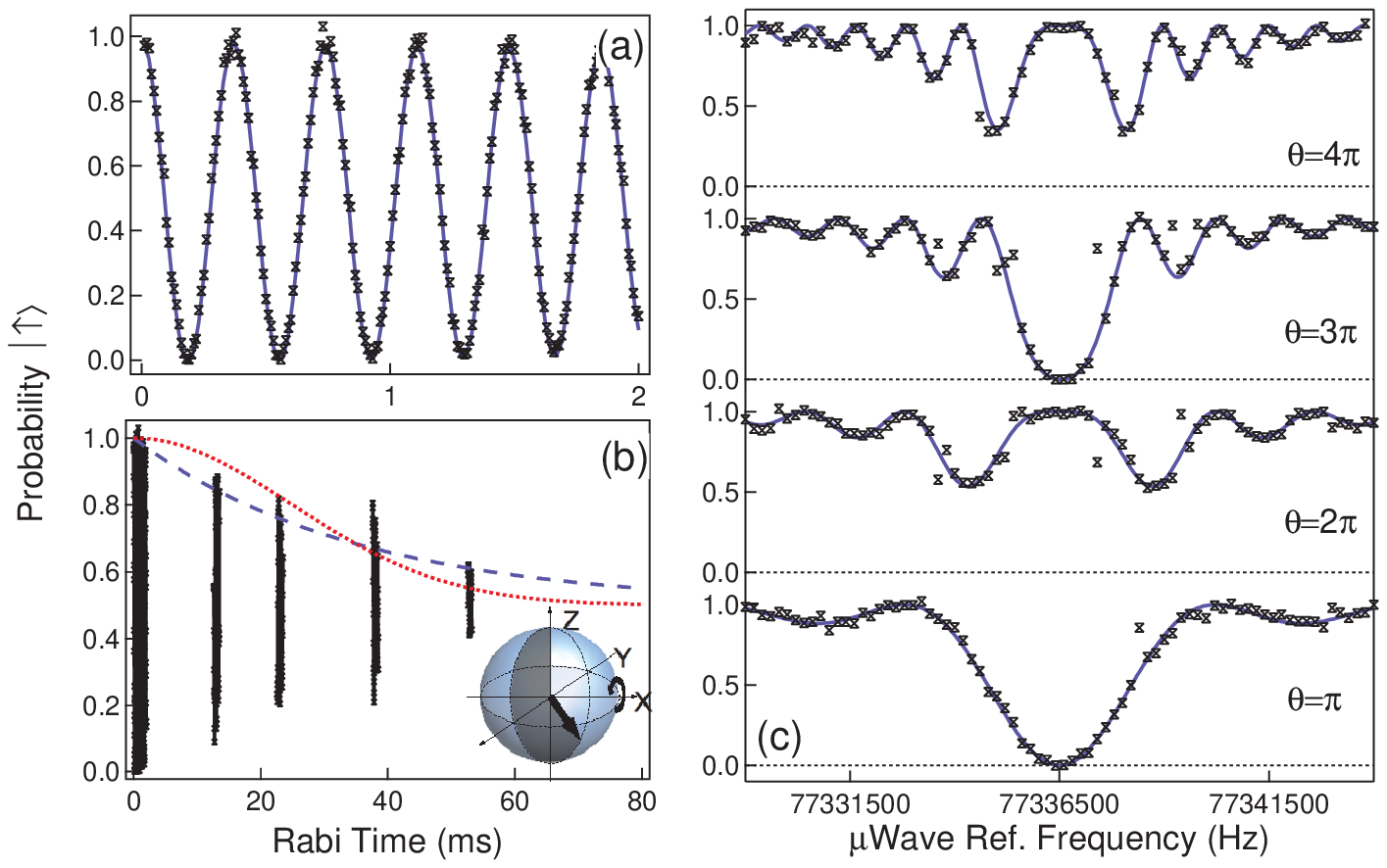, width=12cm}} 
\vspace*{13pt}
\fcaption{\label{fig:Rabi} Driven qubit rotations using 124 GHz microwave system. a) Rabi flopping with $\tau_{\pi}=185\;\mu$s. b) Decay of Rabi flopping contrast for long driving times.  Fringe decay is not well approximated by either a simple exponential (dashed) or a Gaussian (dotted) decay, suggesting the possibility of a complex interplay of processes leading to loss of contrast.  Inset) Graphical depiction of direction of driven rotations on the Bloch sphere.  c) Rabi lineshapes for various rotation angles, determined by the rotation angle $\theta$ ($\theta=\pi$ for $t_{p}=\tau_{\pi}$), as a function of the DDS reference frequency for the 124 GHz microwave system.  Solid lines are a fit to the Rabi lineshape (see text) with a single free parameter - the center frequency, $\omega_{0}$.}
\end{figure}

\noindent  We drive transitions between the electron-spin-flip qubit states using resonant microwave radiation focused on the ions as described in the previous section.  In these experiments we apply a single microwave pulse with variable duration, $t_{p}$, and measure the population in $\left|\uparrow\right\rangle$ as a function of pulse length.  In Fig.~\ref{fig:Rabi}a we demonstrate Rabi flopping with contrast $\sim$99.85~$\%$ (measured for $t_{p}=\tau_{\pi}$) \cite{Ramsey1956, Wineland1998} with a $\pi$-pulse time,  $\tau_{\pi}\approx185$ $\mu$s.  The value of $\tau_{\pi}$ is tunable from roughly 90~$\mu$s to 700~$\mu$s using a manually adjusted attenuator, and is extracted from fits to on-resonance Rabi-flopping curves. The measured contrast arises from a convolution of measurement and operational infidelities.  Typical experimental conditions set $\tau_{\pi}\approx200$ $\mu$s in order to simultaneously maximize flopping contrast and PLL stability.  For shorter values of $\tau_{\pi}$, reflections from the microwave switch mandate manual adjustment of a phase shifter (to mitigate reflections) on an hourly timescale.  The phase of the DDS PLL reference may be changed phase-coherently by 90 degrees in order to affect $\sigma_{X}$ and $\sigma_{Y}$ rotations, where $\sigma_{i}$ is a Pauli matrix.  However, the absolute phase of the DDS reference is insignificant in a single-pulse experiment.

The contrast of Rabi flopping curves can be seen to decay by $\sim$$1/e$ on a timescale of 30-40~ms, as in Fig.~\ref{fig:Rabi}b.  The form of the decay is not well-approximated by either a simple exponential or a Gaussian, suggesting the loss in contrast is due to a complex interplay between multiple processes including long-term phase and amplitude instability of the microwave system \cite{Wineland1998}.  The integrated phase noise of the microwave carrier described earlier produces a phase deviation of a few tenths of a degree over a timescale of a few milliseconds, suggesting phase instability is significant, but not necessarily the only source of the observed decay in Rabi oscillations.  As we will describe in Sec.~\ref{sec:QCoherence}, we believe that the frequency instability of the qubit transition due to fluctuations in the ambient magnetic field dominates expected instabilities in the microwave system.

Rabi lineshapes in the frequency-domain are illustrated for $\theta=\pi$, $2\pi$, $3\pi$, and $4\pi$ rotations in Fig.~\ref{fig:Rabi}c.  These data are obtained by applying a fixed-duration microwave pulse to the qubits, for various values of the $\sim$77 MHz reference output by the DDS.  Fits to these data incorporate the measured $\tau_{\pi}$ and rotation angle, and a single free parameter, $\omega_{0}$, the frequency of the DDS reference on resonance.   The Rabi lineshape \cite{Ramsey1956} is described by
\begin{equation}
1-\left(1+\left(\frac{\omega-\omega_{0}}{2\pi/\tau_{\pi}}\right)^{2}\right)^{-1}\sin^{2}\left[\frac{\theta}{2}\sqrt{1+\left(\frac{\omega-\omega_{0}}{2\pi/\tau_{\pi}}\right)^{2}}\right],
\end{equation}
where $\omega$ is the DDS reference frequency.

The resonance frequency drifts slowly due to flux leakage from our superconducting magnet at a relative rate of $<1\times10^{-8}$ per day.  Additionally, $\omega_{0}$ can occasionally be seen to shift suddenly by up to $\sim$1 kHz, presumably due to the influence of external laboratory equipment in the building.  Careful monitoring of $\omega_{0}$ allows us to maintain long-term microwave frequency stability relative to the qubit transition of $\sim$1$\times10^{-9}$ over any experimental measurement run.

\begin{figure} [htbp]
\vspace*{13pt}
\centerline{\epsfig{file=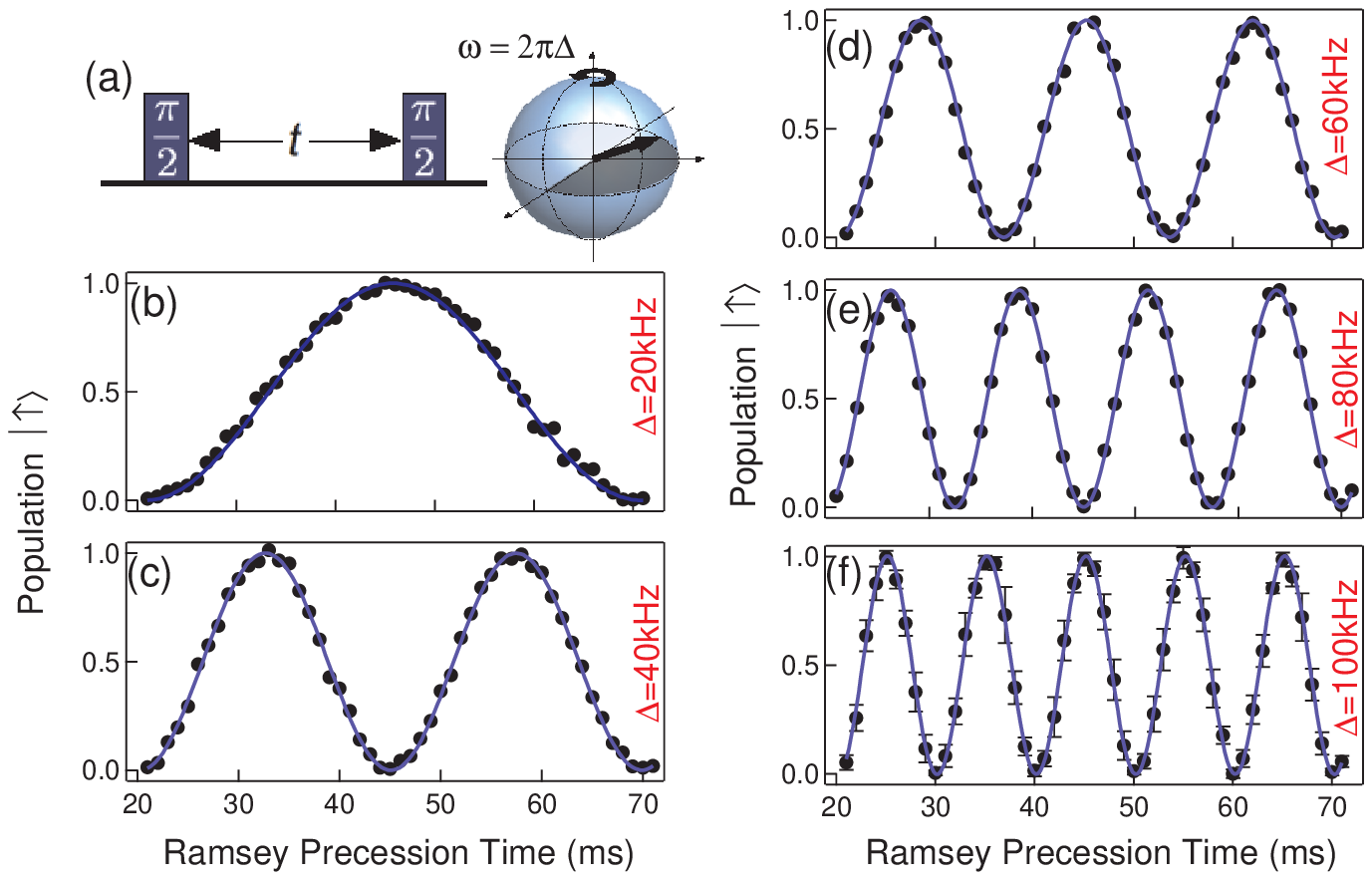, width=12cm}} 
\vspace*{13pt}
\fcaption{\label{fig:Larmor} Controlled Larmor precession.  a) Pulse schematic for Ramsey experiments used to observe Larmor precession.  Microwaves (switch in off state) are detuned from resonance during the interpulse period by $\Delta$, leading to precession at $\omega=2\pi\Delta$. b-f) Qubit state population at the end of the Ramsey experiment as a function of precession time.  Larmor precession frequency increases linearly with $\Delta$ as expected.  Fringes show near unity contrast over measurement range.  The observed 20 $\mu$s offset is due to a delay period during which $\Delta=0$ during free-precession, incorporated for stability of the PLL.  Error bars shown in panel f) depict standard deviation of 20 experiments averaged to produce each point.  Standard deviation is minimized at the extrema of the Ramsey fringes, consistent with quantum projection noise.  Note that actual error per data point is  $1/\sqrt{20}$ of the shown value.}
\end{figure}

\subsection{Controlled Larmor Precession}
\noindent Rotations about the quantization axis in the qubit basis are achieved via detuning of the microwave drive frequency.  We demonstrate controlled Larmor precession using a standard Ramsey interference experiment \cite{Wineland1998}.  We begin a Ramsey experiment by rotating the Bloch vector of the qubits to the equatorial plane via a $\pi/2_{X}$ pulse.  Here the subscript denotes the axis of rotation in the frame rotating at the unperturbed qubit splitting, $\Omega=124$ GHz $+\;\omega_{0}$.  The microwave frequency is then detuned from resonance by an amount $\Delta$, and the qubit is allowed to freely precess.  At the end of the free-precession period another $\pi/2_{X}$ pulse is applied.  The qubit state after the second $\pi/2_{X}$ pulse depends on the phase accumulated by the qubit in the rotating frame, which is manifested as a sinusoidal oscillation in the qubit state population in $\left|\uparrow\right\rangle$ as a function of free-precession time.  The period of the oscillation is $\Delta^{-1}$, and varying $\Delta$ modifies the oscillation period, as shown in Fig.~\ref{fig:Larmor}.  Setting the precession time to $\frac{1}{2}\Delta^{-1}$ while detuned by $\Delta$ performs the Pauli gate $\pi_{Z}$.  The Ramsey fringes measured for $\Delta=100$ kHz are plotted with error bars corresponding to the standard deviation of the $N$ measurements averaged together for each data point (error bars are not normalized by $\sqrt{N}$).  We observe that the standard deviation oscillates as a sinusoid with twice the frequency of the Ramsey fringes, and a $\pi/2$ phase shift.  This yields maximum values of the standard deviation for states which have a 50~$\%$ probability of being projected to $\left|\uparrow\right\rangle$ or $\left|\downarrow\right\rangle$, and minimum deviations for states projecting with unit probability to either of the basis states.  The observed periodicity and normalized values of the standard deviation are consistent with previous studies of quantum projection noise \cite{itaw93}.

\subsection{Randomized Gate Benchmarking}
\noindent We characterize the operational fidelity in our system using a simplification of the randomized benchmarking procedure introduced and demonstrated in reference \cite{Knill2008} and subsequently employed in \cite{Laflamme2008, Schoelkopf2009}.  This procedure provides a simple mechanism to characterize the average fidelity of computationally relevant Clifford operators without the onus of performing full process  tomography.  Clifford operations implemented as driven $\pi/2$ gates about random axes are performed sequentially, with interspersed Pauli $\pi$ rotations performed about arbitrary axes

\begin{figure} [htbp]
\vspace*{13pt}
\centerline{\epsfig{file=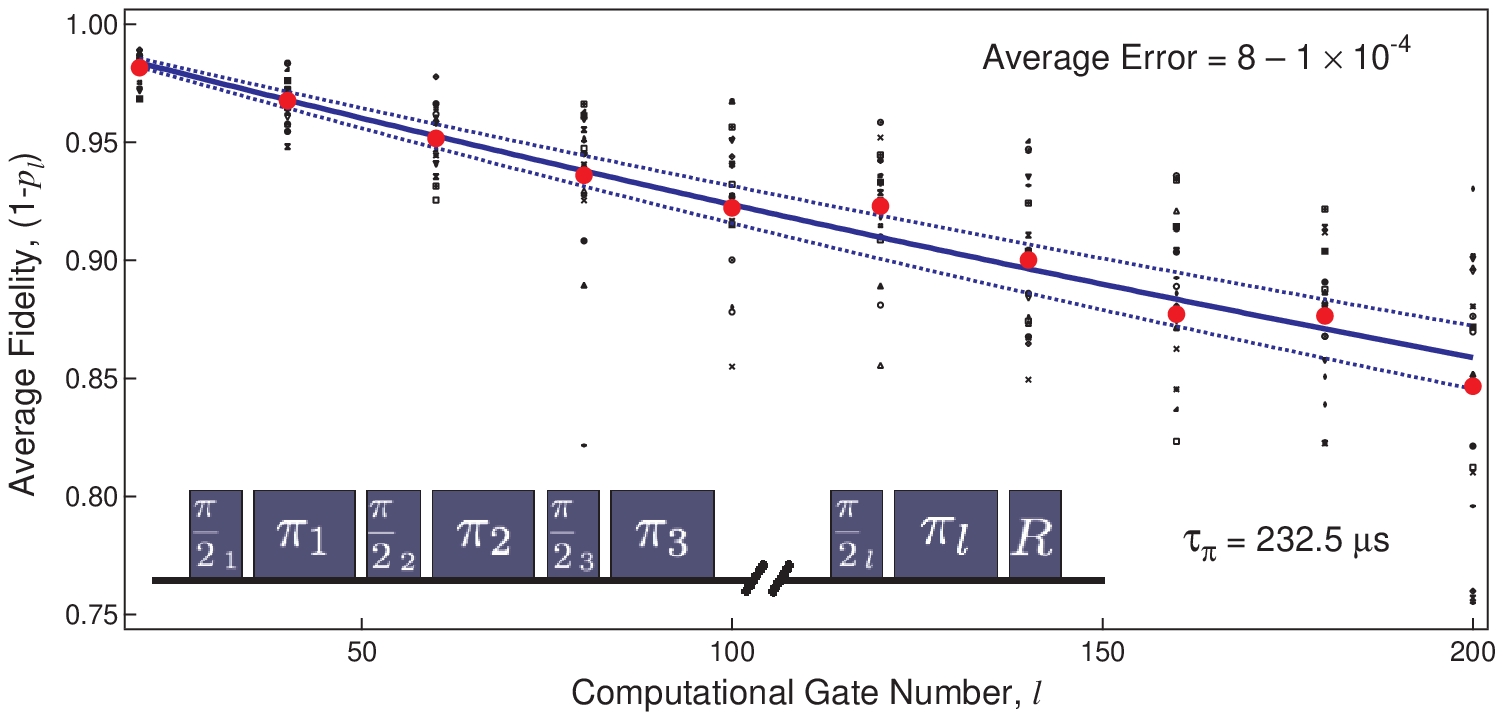, width=12cm}} 
\vspace*{13pt}
\fcaption{\label{fig:Benchmark} Randomized benchmarking experiments using driven $X$ and $Y$ operations.  Randomized sequence of $l$ computational gates constructed as depicted in inset.  Final rotation is deterministic and rotates qubit to dark state as described in the text.  Small markers represent average fidelity of an individual randomized sequence of $l$ gates, with each point consisting of 20 averaged measurements of the same sequence.  Average fidelity defined as ($1-P_{\left|\uparrow\right\rangle}$).  Large markers show average across $k=20$ randomizations for each value of $l$.  Solid line indicates weighted exponential fit to the data, and dashed lines indicate 68~$\%$ confidence level of the fit. (Inset) schematic depiction of the pulse sequence employed.  All gates are separated by a 5~$\mu$s delay during which microwave drive is turned off to allow the PLL to track phase jumps in the external reference signal}
\vspace*{13pt}
\end{figure}

\noindent in order to randomly shift the reference frame.  A single computational gate in this construction is therefore composed of both a Clifford $\pi/2$ and a Pauli $\pi$ operation, each about a randomized axis, as depicted in the inset to Fig.~\ref{fig:Benchmark}.  $k$ runs of length $l$ random computational gates are performed, and the resulting fidelities are averaged for each value of $l$.   For independent, depolarizing errors, the error probability for an $l$-gate sequence, $p_{l}$, increases exponentially with $l$. Fitting an exponential decay as a function of $l$ to the average of the $k$ randomizations thus provides a measure of the error per computational gate.

We assemble a sequence of randomized computational gates and apply them to an initial qubit state $\left|\uparrow\right\rangle$.  $X$ operations are driven on resonance with $\tau_{\pi}=232.5\;\mu$s; driven $Y$ operations are realized by shifting the phase of the $\sim$77 MHz reference to the PLL provided by a DDS by 90 degrees, using TTL inputs.  Each operation is separated by a delay of 5 $\mu$s to ensure that the PLL tracks the phase changes of the DDS reference.  Software keeps track of the expected outcome at the end of the computational sequence and applies an appropriate rotation to the Bloch vector such that in the absence of any error the state is rotated to $\left|\downarrow\right\rangle$ for measurement.  This is accomplished by generating an appropriate rotation matrix for each Clifford and Pauli operation, performing matrix multiplication to simulate the string of computational gates, and generating an appropriate final rotation based on the expected state at the end of the computation.  This implementation is an approximation to the prescription published in Ref. \cite{Knill2008} which omits $\sigma_{Z}$ operations as they simply require an arbitrary change of basis and can have much lower error rates than actual driven operations.   Additionally, we perform all rotations with uniform sign.  Despite these limitations this construction provides a good approximation to the procedure outlined previously, and focuses on the limiting operational fidelities associated with driven microwave transitions.

In Fig.~\ref{fig:Benchmark} we present results for $k=20$ random gate sequences with up to $l=200$ gates.  We plot the average fidelity as a function of $l$;  the fidelity decays to a value of 0.5 for complete depolarization.  Fitting a simple exponential decay to the average fidelity as a function of $l$ provides the solid line presented in the figure, yielding an average error per computational gate of $8\pm1\times10^{-4}$.  Dashed lines indicate the 68~$\%$ confidence interval for the fit.  These results are among the best reported to date for atomic systems, and the low error per computational step is largely due to the extraordinary stability of driven microwave rotations and the absence of spontaneous emission probability in microwave transitions, in contrast to laser-mediated gates.  We believe the primary limitation on operational fidelity is due both to ambient magnetic field fluctuations and to small inhomogeneities in the magnetic field over the extent of the ion crystal, as will be discussed in Sec.~\ref{subsec:Inhomogeneity}.  Additionally, pulse timing resolution of 50 ns corresponds to a systematic $\pi$-pulse timing infidelity of up to $\sim2\times10^{-4}$.  Driven $\pi$-pulse rotations, however, are sensitive to timing infidelity only to second order.

These measurements are performed under typical experimental conditions, and no extra care has been taken to mitigate suspected sources of error.   Previous experiments in superconducting qubits \cite{Schoelkopf2009} have demonstrated a dependence of $p_{l}$ on $\tau_{\pi}$, and, based on the relevant sources of microwave instability in our system, it should be possible to improve our extracted $p_{l}$ by reducing the length of driven rotations.  However, in our current system, substantially reducing $\tau_{\pi}$ has been observed to produce frequency instabilities in the microwave system at pulse edges.

\section{Qubit Coherence}\label{sec:QCoherence}
In this section we present a series of measurements designed to provide information about the limits of qubit coherence in our experimental system.  We study free induction decay in a Ramsey interference experiment, relaxation in the rotating frame via spin-locking, and the suppression of dephasing by application of dynamical decoupling pulse sequences.  In addition, we present an analysis of the effects of magnetic field inhomogeneities in our experimental apparatus.

\subsection{Free Induction Decay}
\begin{figure} [htbp]
\centerline{\epsfig{file=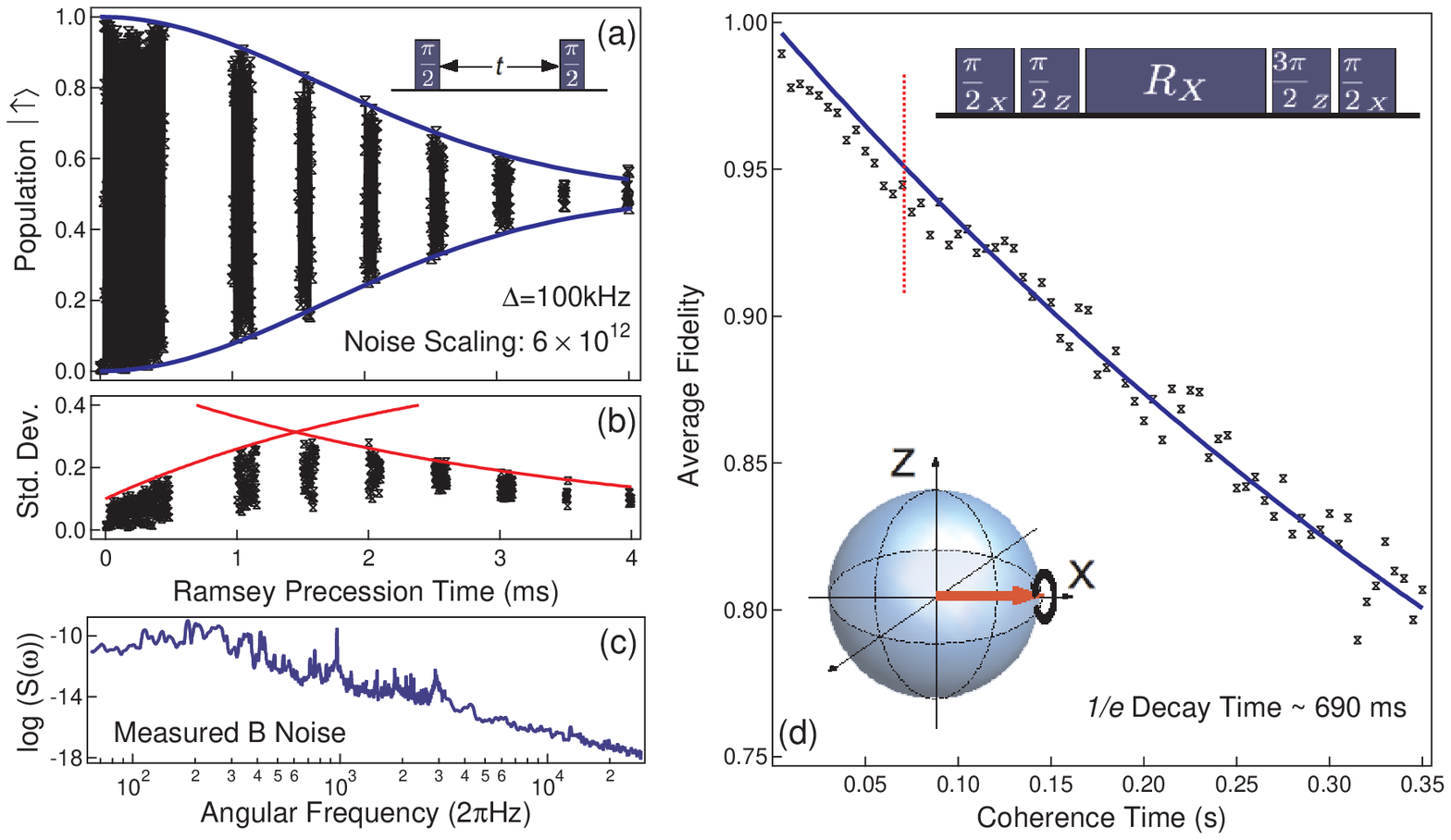, width=12cm}} 
\vspace*{13pt}
\fcaption{\label{fig:Coherence} Studies of qubit coherence a) Decay of Ramsey fringes performed with controlled Larmor precession.  Solid lines are a fit to the envelope of the decay using the coherence integral and filter function as introduced in Eq.~\ref{eq:FF}, and the measured magnetic-field noise presented in panel (c).  Each data point consists of the average of 20 experiments.  b) Standard deviation of averaged data in a).  Standard deviation oscillates consistent with quantum projection noise.  Amplitude of standard deviation grows with increasing time, commensurate with a decrease in fringe contrast.  This indicates the dominance of homogeneous dephasing processes up to free-precession times of $\sim1.5$ ms.  For longer times, magnitude of standard deviation decreases, consistent with the onset of inhomogeneous dephasing across the atomic ensemble.  Solid lines are guides to the eye showing exponential scaling. c) Magnetic field noise measured using a spectrum analyzer connected to a shim coil in our NMR magnet, plotted in arbitrary units on a logarithmic scale. Noise power decays with an approximate $1/\omega^{4}$ scaling to a low-frequency cutoff near 30 Hz.  d) Decay of coherence in a spin-locking experiment fit to a simple exponential, and giving an approximate $T_{1\rho}=688$ ms.  Insets depict pulse sequence and graphical representation of experimental rotation.  $R_{X}$ indicates a rotation about the $X$ axis for a variable time.  Dashed vertical line indicates the approximate experimental duration associated with 200 computational gates in a randomized benchmarking experiment, and corresponds to an error $\sim$3$\times$ lower than that measured via benchmarking. $Z$-rotations enacted by detuning the microwave frequency for a controlled time while microwave switch is off, producing a passive frame change relative to the initial rotational direction.}
\end{figure}

\noindent In addition to demonstrating controlled $\hat{\sigma}_{Z}$ operations, Ramsey free-induction decay measurements provide information about the coherence time of the qubit system \cite{Wineland1998, Vandersypen2004}.  In Fig.~\ref{fig:Coherence}a we show Ramsey fringes obtained using $\Delta=100$ kHz to 4 ms precession time.  Fringe contrast decays as a function of precession time due to a loss of phase coherence between the microwave drive and the qubit system.

We consider the phase coherence of the qubit based on the following analysis.  Starting with a qubit Bloch vector oriented along the $\hat{Y}$-axis in the equatorial plane, we employ a measure of the coherence as $W(t)=\left |\overline{\left\langle\sigma_{Y}(t)\right\rangle}\right |$, where the brackets indicate an expectation value and the overline indicates an ensemble average \cite{Uhrig2007, Cywinski2008, Biercuk2009, BiercukPRA2009}.  In a frame rotating at the qubit precession frequency, and in the absence of dephasing, the state remains oriented along the $\hat{Y}$ axis and $W(t)=1$.  In our system, classical magnetic field fluctuations form the dominant source of noise \cite{Biercuk2009}, and hence provide the limiting mechanism for qubit coherence.

\noindent The longitudinal relaxation time, $T_{1}$ is effectively infinite in this system and will not be considered in our discussion.  A Hamiltonian accounting for the presence of this noise may be written as \cite{Uhrig2007, Uhrig2008, Biercuk2009, BiercukPRA2009}
\begin{equation}\label{eq:Hamiltonian}
\hat{H}=\frac{\hbar}{2}\left[\Omega+\beta(t)\right]\hat{\sigma}_{Z},
\end{equation}
\noindent where $\Omega$ is the unperturbed qubit splitting, and $\beta(t)$ is a time-dependent classical random variable describing the noise in the system.  Under the influence of this Hamiltonian a state originally oriented along $\hat{Y}$ in the rotating frame may be written as
\begin{equation}
\left|\Psi(t)\right\rangle=\frac{1}{\sqrt{2}}\left(e^{-\frac{i}{2}\int_{0}^{t}\beta(t')dt'}\left|\uparrow\right\rangle+e^{\frac{i}{2}\int_{0}^{t}\beta(t')dt'}i\left|\downarrow\right\rangle\right).
\end{equation}
\noindent
The presence of $\beta(t)$ modulates the energy splitting of the Hamiltonian, such that in a frame rotating with angular frequency $\Omega$, the Bloch vector accumulates a random phase shift relative to the $\hat{Y}$-axis at time $t$.  We characterize $\beta(t)$ in the frequency domain by the noise power spectrum $S_{\beta}(\omega)$, defined as the Fourier transform of the two-time correlation function of $\beta(t)$.  The influence of the noise term enters the measure of qubit coherence at time $\tau$ by writing $W(\tau)=e^{-\chi(\tau)}$, where
\begin{equation}\label{eq:FF}
\chi(\tau)=\frac{2}{\pi}\int\limits_{0}^{\infty}\frac{S_{\beta}(\omega)}{\omega^{2}}F(\omega \tau) d \omega.
\end{equation}
In the expression above, $F(\omega\tau)$ is known as a filter-function \cite{Uhrig2007, Cywinski2008}  which encapsulates the experimental conditions under which qubit coherence is measured.  In a Ramsey experiment the filter function takes the form
\begin{equation}
F(\omega\tau)= 4 \sin^{2}(\omega\tau/2),
\end{equation}
such that we expect an approximate Gaussian decay of coherence as a function of the qubit free-precession time, when the small angle approximation is valid.  The solid line in Fig.~\ref{fig:Coherence}a is an envelope function for the decay of qubit coherence based on the construction presented above.  We plot $\frac{1}{2}(1-W(t))$ such that complete loss of phase coherence results in a measured value of $\frac{1}{2}$, commensurate with the decaying contrast of Ramsey fringes.  This function incorporates the filter function appropriate for a Ramsey experiment and the experimental noise spectrum as plotted in Fig.~\ref{fig:Coherence}c.  Magnetic field fluctuations in our superconducting magnet were measured with a solenoid embedded in the magnet system and a spectrum analyzer.  However, the vertical scale of this spectrum depends on the inductance of the solenoid which is not precisely known, and is hence plotted in arbitrary units.  Accordingly, the fitting function contains one free parameter, $\alpha$, which is a multiplicative factor characterizing the strength of the noise.  With this single free parameter we obtain very good agreement between the form of the envelope function and the measured decay of Ramsey fringes, and find a $1/e$ decay time for fringe contrast $T_{2}\approx$~2.4~ms.

The experimental noise spectrum displayed in Fig.~\ref{fig:Coherence}c shows that for angular frequencies beyond $\sim$2,000 rad/s (f$\sim$300 Hz), $S_{\beta}(\omega)$ is reduced by more than four orders of magnitude from its low-frequency peak.  Therefore, for qubit free-precession times $\tau\ll$3 ms, the ambient magnetic field is approximately constant, and the observed decoherence (recall each point represents the results of mutliple experiments that have been averaged together) is due to fluctuations in the magnetic field from experiment to experiment (so-called shot-to-shot variations).  Integrating the measured noise spectrum, and using the scaling parameter extracted from fits to Ramsey oscillations, we obtain an rms shot-to-shot variation in the magnetic field $\delta B/B\sim$1$\times$10$^{-9}$.

\subsection{Effects of Magnetic Field Inhomogeneities}\label{subsec:Inhomogeneity}
Examining the standard deviation of the measurement results for each data point in the Ramsey-fringe curves elucidates the dominant mechanisms of decoherence in the system.  We find that for short times, as described above, the standard deviation oscillates sinusoidally, and in a manner consistent with quantum projection noise \cite{itaw93}.  As we increase the precession time, the sinusoidal variation persists, but the magnitude of the deviation increases commensurate with the beginning of decay in fringe contrast.  As precession time increases further, we observe that the magnitude of the standard deviation begins to decay with what appears to be a simple exponential envelope.  These measurements suggest that for intermediate times ($t<2$ ms), qubit coherence is limited by homogeneous dephasing, where all ion spin vectors are aligned and the system behaves as an effective single spin.  However, as the free-precession time is further increased, the effects of sample inhomogeneities become apparent and will eventually dominate homogeneous dephasing.

In our system, inhomogeneity arises from axial and radial gradients in our quantizing magnetic field.  We characterized the degree of field inhomogeneity by measuring the electron spin-flip transition frequency with a small cloud of ions centered at different axial and radial positions relative to the fixed trap electrodes.  The cloud was moved axially by adding a voltage $\Delta V$ to one endcap and subtracting it from the other.  Radial shifts were achieved by applying asymmetric voltages to the azimuthally segmented electrodes located in the $z$ = 0 plane used to create the rotating wall potential.  At each location of the cloud a resonance profile like that displayed for $\theta=\pi$ in Fig. ~\ref{fig:Rabi}c was taken, and the electron spin-flip frequency was extracted from a fit to the measured Rabi lineshape.


From these measurements we determined an axial gradient of 9.5 kHz/mm (3.4$\times$10$^{-4}$ mT/mm), a maximum gradient transverse
to the z-axis of 3.9 kHz/mm (1.4 $\times$10$^{-4}$ mT/mm), and an average quadratic radial dependence of 3.7 kHz/mm$^{2}$ (1.8
$\times$10$^{-4}$ mT/mm$^{2}$). For single-plane arrays the axial gradient does not produce any inhomogeneous broadening. Occasionally, however, double-plane arrays with 15 $\mu$m spacing were used in experiments, producing a 140 Hz ($\delta B/B\sim10^{-9}$) shift of the qubit resonance frequency between the planes.  The linear gradient transverse to the magnetic field direction should not produce any static frequency shifts due to the averaging effect of the ion array rotation, unlike the quadratic radial magnetic field gradient.  For a typical array radius of 0.3 mm (corresponding to $\sim$600 ions with a 1 kV trap voltage) the quadratic radial field variation produces a 330 Hz ($\delta B/B\sim2.7\times10^{-9}$) shift in the qubit transition frequency between the cloud center and edge.  This shift is small relative to the linewidth of the qubit transition (for our operational $\tau_{\pi}$), but does contribute a small rotation error which may dominate our measured operational infidelity.  We estimate a rotational infidelity of $<1\times10^{-3}$, consistent with the results of randomized benchmarking experiments.

\subsection{Relaxation in the Rotating Frame}
Homogeneous and inhomogeneous dephasing processes may largely be suppressed by the application of spin echo \cite{Hahn50} and other dynamical decoupling pulse sequences, as will be discussed in a later section.  However, it is useful to study the fundamental coherence limits of the qubit in the absence of extra control pulses, as elucidated by a spin-locking measurement \cite{Redfield1955, Haeberlen1976}.  The measurement begins with a $\pi/2_{X}$ pulse which rotates the spin vector to the equatorial plane.  This is followed by a $\pi/2_{Z}$ pulse which rotates the Bloch vector to the $\hat{X}$-axis.  The operation $R_{X}$ is then performed in which the spin state is continuously driven about the $\hat{X}$-axis for a variable time $\gg\tau_{\pi}$.  The experiment concludes with a $3\pi/2_{Z}$ rotation in the equatorial plane, and a $\pi/2_{X}$ rotation to the dark state.  In this experiment the $\sigma_{Z}$ operations are performed by detuning the microwave drive by 100 kHz for a precise time while the microwave switch is off, imparting a controlled Larmor rotation.  We observe that error appears to accumulate in time as a simple exponential with a decay slow compared to the Ramsey free-induction decay time, $T_{2}$.

In a spin-locking experiment, the spin vector ideally remains aligned with the applied microwave field, but over time the alignment of the Bloch vector with the driving field decays.  Early Nuclear Magnetic Resonance (NMR) experiments \cite{Redfield1955} showed that in spin-locking experiments the net magnetization in the equatorial plane would relax over time.  Since the spins were aligned with the strong RF driving field, energy relaxation was required to account for the change in magnetization, hence leading to the concept of the longitudinal relaxation time in the rotating frame, characterized by exponential decay with time constant $T_{1\rho}$.  In these experiments, decay in the rotating frame was attributed to spin-lattice and other similar spin-coupling effects.  However, due to the large ion spacing in our experiments ($\sim$10 $\mu$m) we do not expect such coupling to be significant.  Instead, we attribute the observed decay to the presence of ambient magnetic field fluctuations. This decay is qualitatively different from the effect of magnetic field fluctuations in a Ramsey experiment due to the noncommutative nature of the perturbation with the system Hamiltonian in the present setting.  While the experimental scenario is similar to what is observed in liquid-state NMR, at this time we have not developed a complete theory to model the decay of a spin-locking signal in this environment.


\subsection{Dynamical Decoupling}\label{sec:DD}
\noindent  Dynamical decoupling is a quantum control technique designed to extend qubit coherence and suppress the accumulation of qubit errors by the repeated application of parity-reversing control operations \cite{Viola1998,Viola1999, Zanardi1999,Vitali1999, Byrd2003,Khodjasteh2005, Yao2007, Kuopanportti2008, Gordon2008}.  A simple and familiar version of dynamical decoupling is the Hahn spin echo, widely employed in NMR, electron spin resonance (ESR), and quantum information.  In a frame co-rotating with the unperturbed qubit splitting, $\Omega$, a freely precessing qubit in the presence of environmental noise accumulates  a random phase between the qubit basis states, and hence experiences  dephasing relative to the observer.  However, incorporating a $\pi$ pulse halfway through a well-defined free-precession period approximately time-reverses the accumulation of phase.  So long as fluctuations in the external noise are slow relative to the length of the free-precession period, the effect of the noise is nearly perfectly cancelled at the end of the free-precession period.  This technique was originally designed to mitigate the effects of inhomogeneities in NMR experiments \cite{Hahn50}.  However, the description applied above is also appropriate in the case of a single spin.  Adding $\pi$ pulses to the sequence leads to an extension of the coherence time which scales linearly with the number of pulses, $n$ \cite{Haeberlen1976}.

Considerable research effort has focused on understanding the capabilities of dynamical decoupling pulse sequences for suppressing the effects of both dephasing and longitudinal relaxation, and these techniques are often suggested as candidate error suppression strategies allowing for the realization of fault-tolerant quantum computation.  Recently, it has been discovered that adjusting the relative positions of $\pi$ pulses within a multipulse sequence can improve the noise suppressing capabilities of a given dynamical decoupling pulse sequence by orders of magnitude in particular noise environments \cite{Uhrig2007, Lee2008, Biercuk2009, BiercukPRA2009}. While dynamical decoupling may be employed to mitigate general decoherence processes, we will focus exclusively on dephasing processes in the following discussion.

The assumption that environmental noise is constant over the duration of the sequence is, in general, insufficient, and we must consider how a dynamical decoupling pulse sequence suppresses noise characterized by an arbitrary power spectral density, $S_{\beta}(\omega)$. We again consider a qubit in a noisy environment whose time evolution is given by the Hamiltonian written in Eq.~\ref{eq:Hamiltonian}.  As introduced in Eq.~\ref{eq:FF}, we may write a filter function describing the influence of a dynamical decoupling pulse sequence on qubit coherence \cite{Uhrig2007, Uhrig2008, Cywinski2008}.  For any $n$-pulse sequence one may write a time-domain filter function, $y_{n}(t)$, with values $\pm1$, alternating between each interpulse free-precession period.  The time-domain filter function accounts for phase accumulation under the influence of external noise, with the accumulation being effectively time-reversed by application of $\pi_{X}$ pulses (yielding the alternating sign changes).  We account for nonzero $\tau_{\pi}$ by inserting delays with this length and value zero into $y_{n}(t)$ between free-precession periods.  Moving to the frequency domain we may then write $F(\omega\tau)=|\tilde{y}_{n}(\omega\tau)|^{2}$, where $\tilde{y}_{n}(\omega\tau)$ is the Fourier transform of the time-domain filter function, and $\tau$ is the total sequence length.

Following this construction, for an arbitrary $n$-pulse sequence with length $\tau$ we may write the filter function in the frequency domain
\begin{equation}\label{eq:DDFF}
F(\omega\tau)=|\tilde{y}_n(\omega\tau)|^{2}
=\left|1+(-1)^{n+1}e^{i\omega\tau}+2\sum\limits_{j=1}^n(-1)^je^{i\delta_j\omega\tau}\cos{\left(\omega\tau_{\pi}/2\right)}\right|^{2},
\end{equation}
where $\delta_{j}\tau$ is the time of the center of the $j^{\rm th}$ $\pi_{X}$ pulse, and $\tau$ is the sum of the total free-precession time and $\pi$-pulse times.  It is thus apparent that modifying the relative pulse locations for arbitrary sequence length leads to modification of the filter function.  As $F(\omega\tau)$ enters the coherence integral, $\chi(\tau)$, as a multiplicative factor, we see that the spectral overlap between $S_{\beta}(\omega)$ and the filter function determines the coherence of the qubit.  Modifying the values of $\delta_{j}$ therefore provides a mechanism by which the noise suppressing properties of a dynamical decoupling pulse sequence can be tailored to improve qubit coherence.  Experimental validation of this construction and technique was provided in \cite{Biercuk2009, BiercukPRA2009}.

\begin{figure} [htbp]
\centerline{\epsfig{file=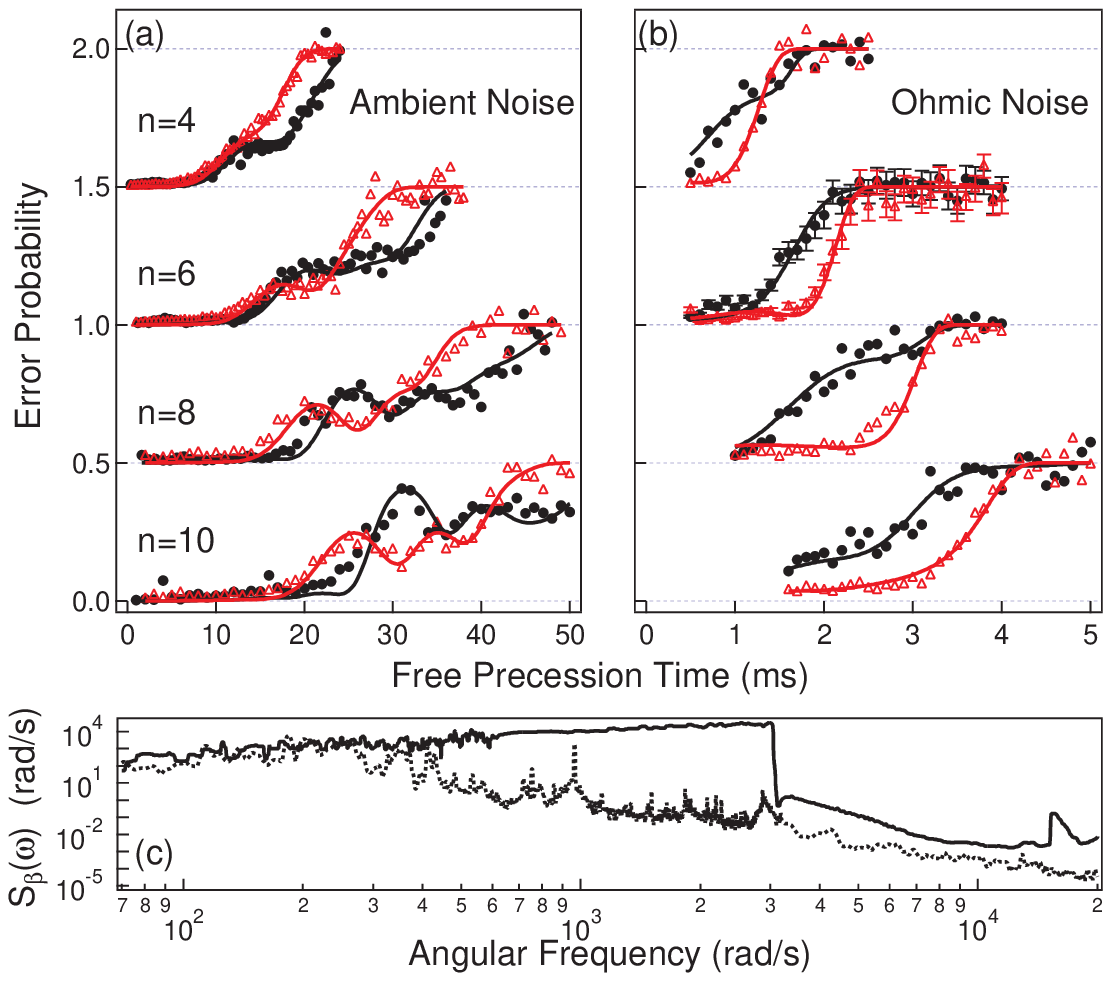, width=9cm}} 
\vspace*{13pt}
\fcaption{\label{fig:DD} Coherence preservation via dynamical decoupling.  (a) CPMG (solid markers) and UDD (open markers) sequence performance in ambient noise for various pulse numbers in a Ramsey interferometer, as a function of total free precession time. Qubits are rotated to $\left|\downarrow\right\rangle$ at the end of an experiment in the absence of dephasing.  The vertical axis corresponds directly to a normalized fluorescence count rate, and hence the probability of completing the experiment in $\left|\uparrow\right\rangle$ due to dephasing.  A value of 0.5 corresponds to complete dephasing.  Here, $\tau_{\pi}=185\;\mu$s.   Curves offset vertically for clarity. Solid lines represent theoretical fits using the appropriate filter function construction as described in Eq.~\ref{eq:DDFF}.  UDD and CPMG perform similarly for all $n$ in the presence of ambient noise with a soft high-frequency cutoff.  (b)  Similar data taken with an artificially engineered noise environment, created using the noise generation system depicted schematically in Fig.~\ref{fig:Electronics}b.  Spectrum is Ohmic ($S_{\beta}(\omega)\propto\omega$) with a sharp high-frequency cutoff at 500 Hz.  In this regime UDD outperforms CPMG, providing significant phase-error suppression over a wide range of free-precession times.  Error bars on curve for $n=6$ derived from normalized standard deviation of averaged experiments and demonstrate statistical significance of differences in sequence performance. (c) Ambient (dotted line) and Artificial Ohmic (solid line) noise power spectra measured directly, and plotted in real units.  Ambient spectrum measured using shim coil in NMR magnet and normalized using a fit to contrast decay of Ramsey fringes (Fig.~\ref{fig:Coherence}(a). Ohmic spectrum derived from measured frequency noise on PLL reference carrier frequency, acquired using a commercial phase-noise detection system.  For details see  \cite{Biercuk2009, BiercukPRA2009}. }
\vspace*{13pt}
\vspace*{13pt}
\end{figure}

We present data studying the effects of two distinct pulse sequences as will be described in order below: CPMG, after Carr Purcell Meiboom Gill, and UDD, or Uhrig Dynamical Decoupling.  CPMG is an extension of the Hahn spin echo to a multipulse form, incorporating evenly spaced $\pi$ pulses about an axis rotated 90 degrees from the direction imparting the initial $\pi/2_{X}$.  The UDD sequence is based on Uhrig's discovery \cite{Uhrig2007} that for an $n$-pulse sequence, enforcing certain constraints on the filter function yielded a sequence which improved the error suppression characteristics of the sequence in the so-called ``high-fidelity'' regime (in which quantum computations are likely to be performed) by orders of magnitude relative to other known sequences.  UDD does a particularly good job at suppressing errors associated with noise power spectra that have sharp high-frequency cutoffs \cite{Uhrig2007, Uhrig2008, Uhrig2008_2, Lee2008, Cywinski2008, Yang2008, Biercuk2009, BiercukPRA2009}.

For these experiments we employ our noise-engineering capabilities and test the error suppressing properties of CPMG and UDD in a variety of experimentally relevant noise environments, as demonstrated originally in \cite{Biercuk2009, BiercukPRA2009}.  In Fig.~\ref{fig:DD} we show experimental results for two different noise environments as well as theoretical fits to our data based on the measure of qubit coherence and the filter function introduced in Eqs.~\ref{eq:FF},~\ref{eq:DDFF}.  These experiments are conducted in a manner similar to the Ramsey coherence measurements described in an earlier section, where the free-precession period is replaced with a sequence of $\pi$ pulses separated by free-precession periods.  We demonstrate that it is possible to both extend the qubit coherence time by increasing $n$, but also that the spectral character of $S_{\beta}(\omega)$ impacts sequence performance as highlighted in Fig.~\ref{fig:DD}a-b.  Theoretical fits show good quantitative agreement with the data and incorporate a single free parameter --- $\alpha$, the strength of the noise.  In these fits, a variety of observed nonlinear features are accurately reproduced by theoretical expressions incorporating the actual $S_{\beta}(\omega)$.  For example, in the ambient spectrum, a sharp feature near 153 Hz is fully responsible for a rounded plateau appearing in the appropriate coherence curves at intermediate times.

Further studies presented in references \cite{Biercuk2009, Uys2009} also demonstrated our ability to develop new suites of locally optimized dynamical decoupling pulse sequences which provide superior error suppression relative to all other known pulse sequences.  These experiments validate a substantial body of literature in the quantum information community and show the versatility of our experimental system for the development of novel and powerful quantum control techniques.

\begin{figure} [htbp]
\centerline{\epsfig{file=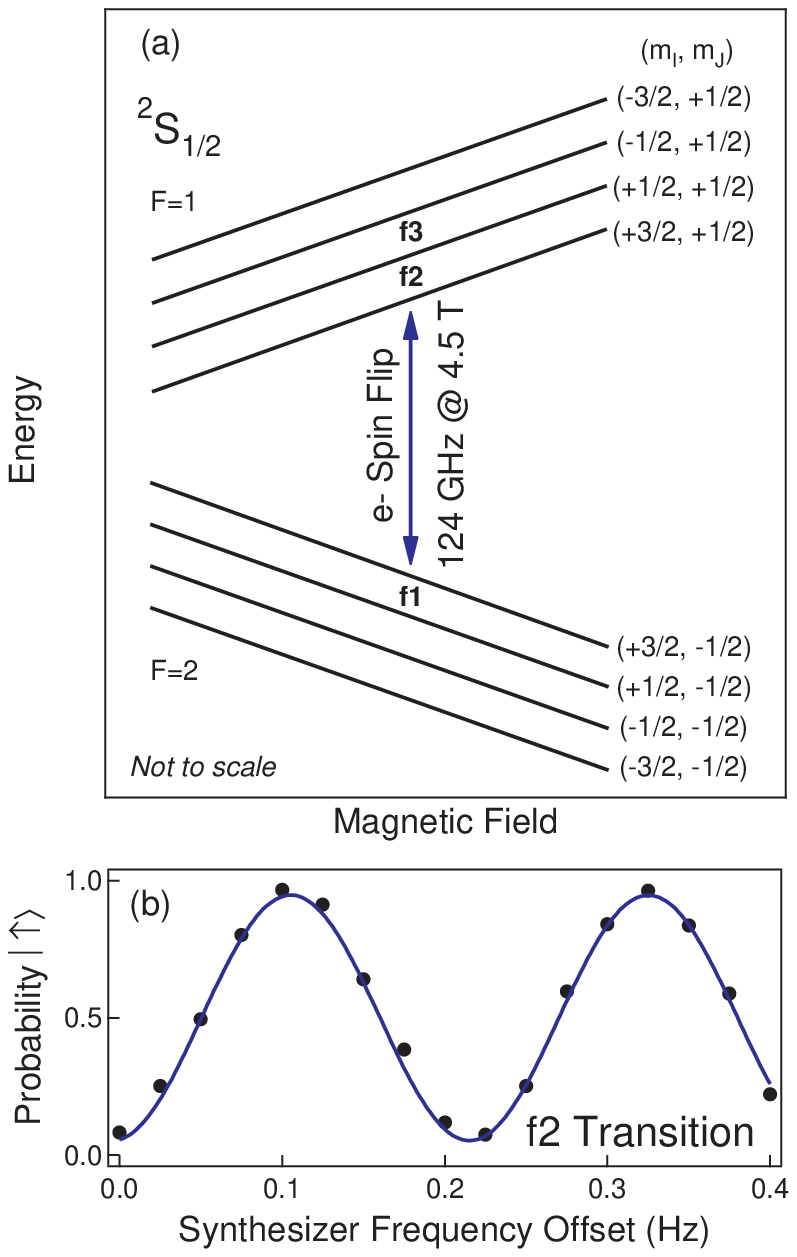, width=6cm}} 
\vspace*{13pt}
\fcaption{\label{fig:Nuclear} Coherent control over a nuclear spin flip in the ground-state manifold of $^{9}$Be$^{+}$. (a) Ground-state energy-level evolution with applied magnetic field. Nuclear spin flip transitions $\textsf{f1}$, $\textsf{f2}$, and $\textsf{f3}$ designated in the figure along with the corresponding ($m_{I}, m_{J}$) quantum numbers.  $\textsf{f1}\sim340$~MHz, $\textsf{f2}\sim288$~MHz, $\textsf{f3}\sim 286$~MHz.  Diagram not to scale. (b) Ramsey fringes measured on the $\textsf{f2}$ nuclear-spin-flip transition as a function of RF frequency offset from 288,172,932.220 Hz.  Frequency tuning range of 0.4 Hz is within bandwidth of $\pi/2$ pulses.  As one state corresponds to the bright state ($\left|\uparrow\right\rangle$) in the electron-spin-flip manifold, we directly probe state population through the measurement of resonance fluorescence.  Total free precession time is 4 s, giving fringe contrast $\sim$90~$\%$.  Microwaves are held at a fixed frequency for the duration of the experiment and the RF switch provides $>$60 dB attenuation in the off-state. }
\vspace*{13pt}
\vspace*{13pt}
\end{figure}

\section{Nuclear Spin Manipulation}\label{subsec:Nuclear}
\noindent Experimental demonstrations discussed thus far have focused on high-fidelity quantum control using a 124 GHz electron-spin-flip transition.  However, this is not the only two-level manifold capable of serving as a qubit in this atomic system. $^{9}$Be$^{+}$ has nuclear spin $I=3/2$ which means that there are three nuclear-spin-flip transitions for each electronic spin state (see Fig.~\ref{fig:Nuclear}).  A state associated with any of these nuclear-spin-flip transitions can be coupled to the optically bright state ($\left|\uparrow\right\rangle$) for the purposes of detection using multiple rf or microwave pulses.  Three of these transitions,
\begin{eqnarray}
\textsf{f1}: \left|\downarrow\right\rangle\equiv\left|F=2, m_{I}=+3/2, m_{J}=-1/2\right\rangle\leftrightarrow\left|F=2, m_{I}=+1/2, m_{J}=-1/2\right\rangle\nonumber\\
\textsf{f2}: \left|\uparrow\right\rangle\equiv\left|F=2, m_{I}=+3/2, m_{J}=+1/2\right\rangle\leftrightarrow\left|F=1, m_{I}=+1/2, m_{J}=+1/2\right\rangle\\
\textsf{f3}: \left|F=2, m_{I}=+1/2, m_{J}=+1/2\right\rangle\leftrightarrow\left|F=1, m_{I}=-1/2, m_{J}=+1/2\right\rangle\nonumber
\end{eqnarray}
at $\sim$288 MHz, $\sim$340 MHz, and $\sim$286 MHz respectively (at 4.5 T) ~\cite{Shiga2009}, are either directly coupled, or may be coupled using a single pulse, to the optically bright electron-spin-flip manifold,  allowing for high-fidelity state initialization and projective measurement through appropriate mapping between the manifolds.

The structure of the energy levels, as represented in Fig.~\ref{fig:Nuclear}a provides these states with low-order immunity to magnetic field fluctuations relative to the main electron-spin-flip qubit transition, making them attractive for long-term information storage, and similar to those used in previous experiments on high-stability frequency clocks~\cite{Bollinger1991}.  The transitions $\textsf{f1}$, $\textsf{f2}$, and $\textsf{f3}$ have been driven using RF fields, and we show a Ramsey experiment on the $\textsf{f2}$ transition ($\textsf{f1}$ and $\textsf{f3}$ not shown) with high contrast for free-precession periods of four seconds.  These experiments employ the RF system depicted schematically in Fig.~\ref{fig:Electronics}c, as well as a loop antenna proximal to the vacuum envelope of the trap.

Unfortunately, the nuclear spin states are weakly coupled to driving fields because of the relatively small value of $g_{I}$, requiring large RF power to drive rotations at speeds comparable to the electron-spin-flip transition.  In Ramsey experiments we realize $\pi/2$ times near 0.5 s, making these operations unreasonably slow relative to the speed of rotations available on the electron-spin-flip qubit.  However, we believe that careful engineering of our RF antenna system will allow the realization of nuclear spin-flip times well below 100 ms.

\section{Future Outlook and Conclusions}\label{sec:Future}
\noindent We have presented a set of experiments demonstrating high-fidelity quantum control using trapped ions in a Penning trap.  Beyond what has been presented in this manuscript and in references \cite{Biercuk2009, BiercukPRA2009, Uys2009}, considerable opportunities exist for exploiting the extraordinary capabilities of this system in future quantum control and quantum information experiments.  In this section we provide a brief discussion of future experimental avenues of interest to us.

\subsection{Improved Operational Fidelity}
Effects due to magnetic field inhomogeneity are experimentally observable (see Fig.~\ref{fig:Coherence}), although they could be minimized by working with small ion arrays.  The measured magnetic field gradients are an order of magnitude larger than those anticipated from the specifications issued by the magnet manufacturer.  This could be accounted for by the fact that the center of our trap is located $\sim$5 mm below the center of the magnet shim coils, and shimming the magnet at this point could have produced a much smaller region of field homogeneity.  Significantly smaller field gradients might be achieved by moving our trap and re-shimming the magnet.  This could enable significantly larger arrays to be used in future studies, improving operational and measurement fidelity.

Of primary interest is the ability to augment the capabilities and performance of our microwave system.  At present our experiments use square microwave pulses implemented via TTL control over a p-i-n diode microwave switch.  Many interesting quantum control experiments may be performed if we have greater control, in real time, over the amplitude of the microwave driving field \cite{Uhrig2008_3, Kuopanportti2008, Gordon2008}.  Accordingly, we are interested in the possibility of using a voltage-controlled attenuator to allow pulse shaping, and even the possibility of new dynamical decoupling techniques that use continuously varied drive fields rather than discrete pulses.

Operational fidelity may be significantly improved by reducing the minimum achievable $\tau_{\pi}$, currently $\sim90\;\mu$s.  This is limited by the power output of the Gunn diode and our ability to focus the microwaves on the ion crystal.  At present our microwave beam waist is $\sim$7 mm at the ion crystal, as determined by the teflon lens we employ.  We have designed and fabricated a new teflon lens that should tighten the microwave beam waist by a factor of two, and hence reduce the area by a factor of four.  The  value of $\tau_{\pi}$ scales linearly with the amplitude of the microwave field incident on the ions, and hence a $4\times$ improvement in power density will be translated to a twofold reduction in $\tau_{\pi}$.  It may also be possible to achieve up to a $5\times$ increase in microwave power generation by replacing the Gunn diode oscillator used in our system.   Further, the requirements of our microwave phase-lock-loop limit us to launching only approximately 10 $\%$ of the Gunn diode's output from the Horn antenna, a limitation that may be lifted by re-engineering our microwave PLL.  Overall, we believe it is possible to achieve more than a 10$\times$ reduction in $\tau_{\pi}$.   Finally, if $\tau_{\pi}$ becomes a critical parameter in future experiments, reducing the magnetic field will allow access to a wide range of commercial microwave components in the W-band near 90 GHz, including higher-ouput Gunn diodes.  In parallel, timing infidelity issues may be mitigated by employing pulse generators with clock frequencies higher than the 20 MHz used through most of our experiments.

The magnitude of integrated shot-to-shot magnetic field variations represents an infidelity of approximately $1\times10^{-3}$ for $\tau_{\pi}\approx200$ $\mu$s.  Increasing the length of the magnetic field vector due to the driving field in the rotating frame by 10$\times$ will result in a 100$\times$ improvement in operational fidelity for fixed magnetic field noise.  This will allow us to realize operational fidelities well below $1\times10^{-4}$, appropriate for tests in the ``fault-tolerant'' regime.

\subsection{Improved Detection Fidelity}
Our present experimental set-up uses an $f/5$ solid angle collection of the ion fluorescence scattered perpendicular to the magnetic field. The light collection is limited primarily by the size of the hole in the ring electrode which provides optical access.  $f/2$ light collection may be achieved parallel to the magnetic field, but in the current set-up this access is blocked by the teflon lens used for focusing the 124 GHz microwaves on the ion array. A light collection of better than f/2 (while simultaneously admitting microwaves) should be possible in a new trap, improving our light collection by more than a factor of 6.

In the absence of dark counts or stray laser light scattered off trap electrodes, detection fidelity is typically limited by the mean number of photons scattered by the bright state in the amount of time required for the dark state to be repumped to the bright state by the detection laser.  For the 124 GHz electron spin qubit discussed here, this is the number of photons that are scattered on the cooling transition (see Fig.~\ref{fig:Levels}) before the $\left|\downarrow\right\rangle = (+3/2, -1/2)$ state is repumped to the $\left|\uparrow\right\rangle = (+3/2, +1/2)$ state.  In general, this scattering rate improves with increased magnetic field due to the energy splitting of the excited state in $^{9}$Be$^{+}$.  At 4.5~T the cooling laser is detuned from the repumping transition by 40 GHz.

With an $f/2$ light collection and assuming a 20~$\%$ quantum efficiency we estimate that on average $\sim$20,000 photons can be detected for each bright-state ion before $\left|\downarrow\right\rangle$ is repumped to $\left|\uparrow\right\rangle$.  With an array of a few hundred ions this should permit detection infidelities at the 10$^{-4}$ level, allowing the observation of quantum jumps when most of the ions are in the dark state.

\subsection{Exploitation of the Nuclear Spin}
We have directly driven the RF nuclear spin flip transitions discussed in the last section of this manuscript, but to date we have not employed them as independent qubit manifolds with coherent mapping to and from the electron-spin-flip transition in an algorithmic setting.  We envision both long-term information shelving \cite{Dutt2007}, and potentially even multiqubit algorithms and advanced quantum control procedures which involve multiqubit interaction \cite{Morton2006}.  These capabilities are predicated on our ability to achieve stronger coupling of the nuclear spins to the RF driving field, which will require substantial, but straightforward modifications to the RF antenna system.

\subsection{Single Ion Addressing}
Many quantum information studies can be pursued without addressing individual ions.  Examples include the
dynamical decoupling studies discussed here, and the simulation of quantum systems \cite{Porras2006, Hasegawa2008} through the generation of large-scale entanglement.  More complex experiments, including procedures outlined in \cite{Taylor2007}, may eventually require single-ion addressing.  While challenging, it may be possible to use strobed or co-rotating laser-beams locked to the frequency of the rotating wall potential, similar to the techniques used for single-ion imaging.   We discuss here the prospects for individual ion addressing with well
focussed laser beams directed perpendicular to the ion array.

In Section~\ref{subsec:TrapControl}, we described the stability of the ion crystal in the frame of the rotating wall potential.  The periods between ``slipping'' events --- tens of seconds --- are sufficiently long to make single ion addressing with well focused, rotating laser beams feasible. Crossed acousto-optic modulators, combined with an imaging system, can be used to generate a tightly focussed laser beam which is displaced from, but rotates about, the rotation axis of the planar array.  This technique is similar to that demonstrated with stirred Bose-Einstein condensates ~\cite{paris2000, mit2001}.

Clearly, increasing the length of the time periods where the planar array is locked to the rotating wall improves the feasibility of single ion addressing and motivates further studies like that discussed in \cite{Mitchell2001}.  For instance, the use of a strong rotating quadrupole potential which generates a strong perturbation to the ion array boundary should increase the length of the ``sticking periods''.  This regime may be reached with a minimization of the distance between the ions and the electrodes imparting the rotating wall potential.  It will also be useful to reduce the deleterious effects of the perpendicular laser beam torque.  This may be achieved by appropriate tailoring of the perpendicular laser beam profile;  for example use of a large beam waist with frequency dispersion across the waist that matches the ion Doppler shifts due to the plasma rotation.

Rather than precisely rotating a well focussed laser beam as discussed above, single ion addressing should also be possible with a fixed laser beam and an appropriate center-of-mass (COM) shift in the planar array. Consider an ion located a distance $r_0$ from the center of a single species planar array whose rotation frequency $\omega_r$ is set by a rotating (e.g. quadrupole) potential.  The position of the ion can be parameterized by $r(t) = r_0 e^{-i(\omega_r t + \delta)}$. We can now apply a rotating electric field which drives the COM motion of the ions.  By rotating the field at $\omega_r$ and setting its strength and phase appropriately, the COM displacement can be adjusted to equal $-r(t)$.  In this case the designated ion is fixed while the remaining ions rotate about it, which allows addressing with a fixed laser beam.  This addressing scheme is simpler than rapidly rotating a laser beam, but it faces some limitations.  First, the COM displacement of the array would likely have to be performed slowly relative to the transverse mode frequencies of the array, likely limited by the presence of shear modes.  These modes have not been studied, but their frequencies are reduced by $\mathbf{E}\times\mathbf{B}$ dynamics relative to that of an unmagnetized planar array.  Also, the presence of impurity ions of heavier mass will complicate the COM displacement of the array.  A sufficiently large number of impurity ions of heavier mass will shield the applied electric field and result in no net displacement of the Be$^{+}$ ions~\cite{huap98b}.

\subsection{Conclusion}
In summary, we have demonstrated high-fidelity control over a qubit realized as trapped $^{9}$Be$^{+}$ ion crystals in a Penning trap.  This trap employs static electric and magnetic fields in order to provide charged-particle confinement and yields distinct benefits over RF traps in realizing large ion arrays.  Qubits are defined using an electron-spin-flip transition at 4.5 T, and are driven using a 124 GHz microwave system, providing control over all axes of the Bloch sphere, and long coherence times.  The average infidelity per randomized computational gate is measured through a benchmarking technique to be $8\pm1\times10^{-4}$, providing significant advantages for studies of high-fidelity quantum control.  As an example, we present studies of dynamical decoupling pulse sequences, highlighting our ability to suppress the accumulation of phase errors and artificially engineer the noise environment in the system.  Taken together, these experiments illustrate how ion crystals in a Penning trap are a broadly useful model quantum system.  Future experiments show a path towards higher-fidelity gates, and potential studies of large-scale entanglement and quantum simulation.  Such experiments are challenging technically, but may further improve the value of ion crystals in Penning traps as a fundamental tool for studies of quantum information and quantum control.

\nonumsection{Acknowledgements}
\noindent
The authors thank L. Cywinski, S. Das Sarma, V. V. Dobrovitski, X. Hu, E. Knill, S. Lyon, G. Uhrig, and W. Witzel for useful discussions.  We also thank D. Leibfried and R. Simmonds for their comments on the manuscript, and C. Nelson for technical assistance with phase-noise measurements.  We acknowledge research funding from DARPA, IARPA, and the NIST Quantum Information Program.  M. J. B. acknowledges fellowship support from IARPA and Georgia Tech., and H. U. acknowledges support from CSIR.  This manuscript is a contribution of the US National Institute of Standards and Technology and is not subject to US copyright.

\nonumsection{References}




%

\end{document}